\documentclass[10pt,aps,prd,onecolumn,notitlepage,nofootinbib,floatfix,superscriptaddress]{revtex4-1}

\usepackage{amsmath, amsthm, amssymb, amsfonts, amsbsy, mathrsfs}

\usepackage{xcolor}
\usepackage{calligra,bm}
\usepackage{stmaryrd}
\usepackage{varioref}

\usepackage{graphicx}
\graphicspath{{./images/}}

\RequirePackage[colorlinks,citecolor=blue,urlcolor=magenta,linkcolor=blue]{hyperref} 
\hypersetup{pdfstartview=FitH,pdfhighlight=/O}

\labelformat{section}{Section #1} 
\labelformat{subsection}{Section #1} 
\labelformat{subsubsection}{Section #1}
\labelformat{subsubsubsection}{Section #1}
\labelformat{equation}{Eq.~(#1)} 
\labelformat{figure}{Fig.~#1} 
\labelformat{subfigure}{Fig.~\thefigure#1} 
\labelformat{table}{Tab.~#1} 
\labelformat{appendix}{Appendix #1}

\begin{document}

\title{Weiss variation for general boundaries}

\author{Justin C. Feng}
\email{justin.feng@tecnico.ulisboa.pt}
\affiliation{CENTRA, Departamento de F{\'i}sica, Instituto Superior T{\'e}cnico – IST, Universidade de Lisboa – UL, Avenida Rovisco Pais 1, 1049 Lisboa, Portugal}

\author{Sumanta Chakraborty}
\email{Corresponding Author}
\email{tpsc@iacs.res.in}
\affiliation{School of Physical Sciences, Indian Association for the Cultivation of Science, Kolkata-700032, India}


%
%

%
%
\begin{abstract}
The Weiss variation of the Einstein-Hilbert action with an appropriate boundary term has been studied for general boundary surfaces; the boundary surfaces can be spacelike, timelike, or null. To achieve this we introduce an auxiliary reference connection and find that the resulting Weiss variation yields the Einstein equations as expected, with additional boundary contributions. Among these boundary contributions, we obtain the dynamical variable and the associated conjugate momentum, irrespective of the spacelike, timelike or, null nature of the boundary surface. We also arrive at the generally non-vanishing covariant generalization of the Einstein energy-momentum pseudotensor. We study this tensor in the Schwarzschild geometry and find that the pseudotensorial ambiguities translate into ambiguities in the choice of coordinates on the reference geometry. Moreover, we show that from the Weiss variation, one can formally derive a gravitational Schr{\"o}dinger equation, which may, despite ambiguities in the definition of the Hamiltonian, be useful as a tool for studying the problem of time in quantum general relativity. Implications have been discussed.
\end{abstract}


\maketitle


%
%

%
%

\section{Introduction}

\begin{figure}[!t]
    \includegraphics[width=0.5\columnwidth]{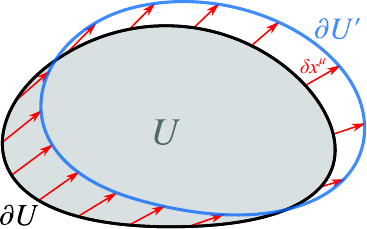}
    \caption{
    Illustration of boundary displacement in the Weiss variation.
    }
\label{fig:Boundary_Sweep}
\end{figure}

For every dynamical variable, there exists an action functional $S$, whose extremization under arbitrary variation of the variable, yields the corresponding classical equations of motion. In addition, the integral of the phase factor $\exp(iS/\hbar)$ over all possible paths leading to a certain configuration of the dynamical variable, provides the wave function $\psi$ of the dynamical variable, such that $|\psi|^{2}$ provides the probability of finding the variable in the desired configuration. Constructing such an action functional, along with a well-behaved variational principle is central to any field theory. For gravity, the situation is more complex, thanks to the equivalence principle, which requires any covariant action functional for gravity to have second order derivatives of the metric. Therefore, one needs to fix both the metric and its normal derivatives at the boundary, rendering the variational principle ill-posed \cite{York:1972sj,Gibbons:1976ue,Charap:1982kn,Dyer:2008hb}. In order to get rid of this problem, suitable boundary terms must be added to the action functional, which depends on the nature of the boundary surfaces \cite{Gibbons:1976ue,Padmanabhan:2014lwa,Parattu2015,Lehner:2016vdi,FengMatzner2018}. Given the issues with the well-posed nature of the variational problem for the gravitational action functional, we wish to explore a modified variational problem, namely that of Weiss variation \cite{Weiss1936,SudarshanCM,MatznerShepleyCM,FengMatzner2018}. In contrast to the standard variational problems, the Weiss variation, on the other hand, is a variation of the action which involves boundary/endpoint variations as well, thereby including displacements of the boundaries in the manifold (for a pictorial demonstration, see \ref{fig:Boundary_Sweep}). To familiarize the reader with the basics of the Weiss variation, we start with a mechanical action of the following form:
\begin{equation} \label{WAC-MechAction}
S[q]:=\int^{t_2}_{t_1} L(q,\dot{q},t) \> dt,
\end{equation}
\noindent the full variation of the action under endpoint displacements may be written as:
\begin{equation} \label{WAC-FvarMechAction}
  \Delta S = \int^{t_2}_{t_1} \mathbb{E} \cdot \delta q \> dt + \biggl[ p \cdot \delta q + L \Delta t \biggr] \biggr |^{t_2}_{t_1},
\end{equation}
\noindent where $\mathbb{E}=0$ are the Euler-Lagrange equations, $\Delta t$ is the infinitesimal endpoint displacement, in this case that of time (evaluated at $t_i$, with $i\in\{1,2\}$), and $\delta q$ represents the variation in the function $q=q(t)$. Ideally, the endpoint terms should be written in terms of the total displacement in $q$, which may be defined as:
\begin{equation} \label{WAC-Deltaq}
  \left. \Delta q \right|_{t_i}:=q(t_i+\Delta t_i)+\delta q(t_i+\Delta t_i)-q(t_i).
\end{equation}
\noindent To first order in the variation, the total displacement differs from $\delta q|_{t_i}$ in the following way:
\begin{equation} \label{WAC-Variationq}
  \left. \Delta q \right|_{t_i}=\left. \delta q + \dot q \, \Delta t \right|_{t_i}.
\end{equation}
\noindent Solving for $\delta q$, one obtains the Weiss form of the variation:
\begin{equation} \label{WAC-WvarMechAction}
\Delta S = \int^{t_2}_{t_1} \mathbb{E} \cdot \delta q \> dt + \biggl[ p \cdot \Delta q - H \> \Delta t \biggr] \biggr |^{t_2}_{t_1},
\end{equation}
\noindent where $H := p \cdot \dot{q} - L$, which one may recognize to be the Hamiltonian of the system. For a mechanical system, one may then identify the conjugate momenta and Hamiltonian as factors appearing in front of the respective endpoint displacements $\Delta q$ and $\Delta t$, respectively. More detailed discussions of the Weiss variation in mechanics may be found in \cite{SudarshanCM,MatznerShepleyCM,FengMatzner2018}.

Along identical lines, for a classical field theory with action $S_\varphi=S_\varphi[\varphi^a]$, the Weiss variation generalizes to:
\begin{align} \label{WAC-WeissVariationFields}
  \Delta S_{\varphi}&=\int_{U} d^4 x \, \mathbb{E}_a \, \delta \varphi^a 
  \nonumber
  \\
  & + \int_{\partial U} d\Sigma_\mu \biggl[ \pi{^\mu}_a \, \Delta \varphi^a
  - (\Theta{^\mu}{_\sigma} \, \xi^\sigma + S^\mu) \, \Delta t
  \biggr],
\end{align}

\noindent where $\mathbb{E}_a=0$ are the field equations, $\pi^\mu_a := \partial L/\partial (\partial_\mu\varphi^a)$ is the polymomentum, $\Theta{^\mu}{_\beta}$ is the canonical energy-momentum tensor, and $S^\mu$ is a term which yields the Belinfante-Rosenfeld tensor \cite{Belinfante1940,*Rosenfeld1940} (as we shall later show in this article). The quantity $\xi^\mu$ is a vector field which characterizes an infinitesimal displacement $\delta x^\mu = \xi^\mu \, \Delta t$ of the boundary $\partial U$ parameterized by the infinitesimal parameter $\Delta t$ (see \ref{fig:Boundary_Sweep}). The quantity $\Delta \varphi^a$ is the difference between the varied field evaluated at the displaced boundary $\partial U^\prime$ and unvaried field evaluated at the boundary $\partial U$. With an appropriate choice of boundary surface $\partial U$, and vector field $\xi^\mu$, one can obtain a Hamiltonian for the field theory without performing an explicit (3+1) split (for such an application, see \cite{Fengetal2019}). Moreover, the Weiss variation generalizes in a straightforward way to higher derivative theories (in the mechanical case, one obtains the Ostrogradsky momenta and Hamiltonian), making the Weiss variation a useful tool for constructing the Hamiltonian formulation of complicated field theories.

In this article, we revisit the Weiss variation in general relativity and extend it to the case of general boundary surfaces. The Weiss variation for the Einstein-Hilbert action with the usual GHY boundary term for spacelike and timelike surfaces was obtained in \cite{FengMatzner2018}. However, the expression obtained in that article does not generalize straightforwardly to null boundaries (for an understanding of the appropriate boundary term associated with null surfaces, see \cite{Chakraborty:2016yna,Parattu:2016trq,Parattu2015}). Moreover, one might seek a formalism that is agnostic to the geometric properties of the boundaries. Keeping these in mind, the approach we explore in this article employs a different choice of the covariant boundary term, which is a generalization of the boundary term used to obtain the ``$\Gamma\Gamma$'' action, incorporating a non-dynamical auxiliary connection which does not contribute to the bulk action. Such an approach is not new --- an action constructed in this manner was proposed in \cite{Rosen1940} (see also \cite{Papapetrou1948}), discussed in \cite{IsenbergNester1980} (with the boundary term employed in a simplified proof of the positive energy theorem \cite{Nester1982}), and studied further in \cite{Lynden-Belletal1995}. See also \cite{Harada2020CIGA} and \cite{BeltranJimenez2019} for recent discussions, and also the recent review \cite{PetrovPitts2020}. Despite this, it is not as widely employed as the GHY boundary term, likely due to ambiguities in the choice of the auxiliary connection. However, the fact that this boundary term is agnostic to the signature of the boundary greatly facilitates the derivation of the Weiss variation for general boundary surfaces.

Concerning our notation for the variations, $\delta$ will refer to the usual first-order variation of functions and functionals, assuming fixed boundary surfaces for the functionals. Variations that include boundary displacements will be indicated with $\Delta$. We employ the conventions of \cite{Carroll}, and employ those of \cite{Parattu2015} for null surfaces, with the exception of the equality $:=$, which indicates a definition in this article. Barred quantities ($\bar\nabla_\mu$ and $\bar R_{\mu \nu}$ for instance) refer to quantities defined with respect to a nondynamical reference geometry, characterized by a connection $\bar{\Gamma}{^\lambda}{_{\mu \nu}}$ or a reference metric $\bar{g}_{\mu \nu}$. The reference connection $\bar{\Gamma}{^\lambda}{_{\mu \nu}}$ is not generally assumed to be a Levi-Civita connection with respect to $\bar{g}_{\mu \nu}$, and instances in which this assumption is made will be indicated explicitly.

In \ref{Weiss_grav}, we discuss the Weiss variation of the Einstein-Hilbert action and the covariant boundary term described above. We show that the boundary terms reduce to the expected forms in the appropriate gauges, and work out in detail the expressions for the conjugate momenta on null and non-null boundary surfaces in \ref{geometry_boundary}. In \ref{grav_em_tensor}, we examine in detail the gravitational energy-momentum tensor (a generalization of the Einstein pseudotensor) obtained from the covariant boundary term and discuss its features. In \ref{weiss_matter}, we include matter fields and obtain the full Weiss variation for matter and general relativity in this formalism. Finally, in \ref{einstein_schrodinger}, we present the derivation of an Einstein-Schr{\"o}dinger equation for vacuum quantum general relativity and discuss its implications. Details of the computations have been presented in the appendices.


%
%

%
%
\section{Weiss variation of the gravitational action}\label{Weiss_grav}

\subsection{Variation of the Einstein-Hilbert action}

We begin by performing the Weiss variation of the Einstein-Hilbert action, i.e., variation including contributions from the boundary displacements. For this purpose, we first spell out the Einstein-Hilbert action itself, which takes the form:
\begin{equation} \label{WAC-EHAction}
 S_{\rm EH}[g^{\mu \nu}] = \frac{1}{2 \kappa}\int_{U} d^4 x \, \sqrt{-g} \, {R},
\end{equation}

\noindent where ${R}$ is the Ricci scalar constructed from the Levi-Civita connection ${\Gamma}{^\lambda}{_{\mu \nu}}$ compatible with the metric $g_{\mu \nu}$. The variation of the Ricci scalar ${R}$ induced by variations in the metric tensor components $g^{\mu \nu}$ is:
\begin{equation}\label{WAC-RicciScalarVariation}
\begin{aligned}
\delta {R}
%
%
%
        &= {\nabla}_\mu \left(\delta{^{\mu \alpha}}{_{\nu \sigma}} \, g^{\sigma \beta} \delta {\Gamma}{^\nu}{_{\alpha \beta}}\right) + {R}_{\mu \nu}\> \delta g^{\mu \nu},
\end{aligned}
\end{equation}

\noindent where $\delta{^{\mu \alpha}}{_{\nu \sigma}}$ is the rank $(2,2)$ generalized Kronecker delta, which is defined as:
\begin{equation} \label{WAC-GKD}
  \begin{aligned}
  \delta{^{\mu \alpha}}{_{\nu \sigma}}
      & := \delta{^\mu}{_\nu} \delta{^\alpha}{_\sigma} - \delta{^\mu}{_\sigma} \delta{^\alpha}{_\nu}.
  \end{aligned}
  \end{equation}

\noindent The variations of the inverse metric and volume element take the respective forms:
\begin{equation} \label{WAC-InverseVolumeElementVariation}
\begin{aligned}
\delta g_{\mu \nu} &= - g_{\sigma \mu} \, g_{\tau \nu} \, \delta g^{\sigma \tau}\\
\delta \sqrt{-g} &=-\frac{1}{2} \sqrt{-g} \> g_{\mu \nu} \> \delta g^{\mu \nu}.
\end{aligned}
\end{equation}

\noindent Allowing for boundary variations, the variation of the action presented in \ref{WAC-EHAction} is then given by:
\begin{equation} \label{WAC-ACFActionVar1}
\begin{aligned}
\Delta S_{\rm EH} =&~\frac{1}{2 \kappa} \biggl\{\int_{U} d^4 x  \sqrt{-g} \biggl[\left({R}_{\mu \nu} - \frac{1}{2} {R} g_{\mu \nu}\right) \delta g^{\mu \nu} \biggr] \\
& \qquad + \int_{\partial U} d\Sigma_\mu \left(\delta{^{\mu \alpha}}{_{\nu \sigma}} \, g^{\sigma \beta} \, \delta {\Gamma}{^\nu}{_{\alpha \beta}} + {R} \, \delta x^\mu \right) \biggr\} ,
\end{aligned}
\end{equation}

\noindent where $\delta x^\mu$ is an infinitesimal boundary displacement, and the surface element is defined as $d \Sigma_\mu := \sqrt{-g} \, d \underline{\Sigma}_\mu$, with:
\begin{equation} \label{WAC-SurfElement}
  d \underline{\Sigma}_\mu := \frac{1}{3!} \epsilon_{\mu \alpha \beta \gamma } \, dx^\alpha \wedge dx^\beta \wedge dx^\gamma,
\end{equation}

\noindent where, $\epsilon_{\mu \alpha \beta \gamma }$ is the totally antisymmetric Levi-Civita symbol with $\epsilon_{0123}=+1$. Setting $\delta x^\sigma = \Delta \lambda \, \xi^\mu$, where $\xi^\mu$ is a vector field and $\delta \lambda$ is an infinitesimal parameter, one may write the change of the Levi-Civita connection as the following:
\begin{equation} \label{WAC-DeltaGam}
\begin{aligned}
\Delta {\Gamma}{^\nu}{_{\alpha \beta}} &= \Delta \lambda \pounds_\xi {\Gamma}{^\nu}{_{\alpha \beta}} + \delta {\Gamma}{^\nu}{_{\alpha \beta}},
\end{aligned}
\end{equation}

\noindent where the Lie derivative of the Levi-Civita connection reads:
\begin{equation} \label{WAC-LieDerivConnection}
  \pounds_\xi {\Gamma}^\gamma{_{\mu \nu}}
  = {\nabla}_\mu {\nabla}_\nu \xi^\gamma + \xi^\sigma {R}{^\gamma}_{\nu \sigma \mu} ,
\end{equation}

\noindent yielding:
\begin{equation} \label{WAC-LieDerivConnectioncont}
  \delta{^{\mu \alpha}}{_{\nu \sigma}} \, g^{\sigma \beta} \, \pounds_\xi {\Gamma}^\nu{_{\alpha \beta}}
  = 2 {R}{^\mu}{_\sigma} \xi^\sigma - j^\mu.
\end{equation}

\noindent Here $j^\mu$ is the Noether-Komar current (for interesting thermodynamic implications of the same, see \cite{Chakraborty:2015hna,Padmanabhan:2013nxa,Chakraborty:2018qew}):
\begin{equation} \label{WAC-KomarCurrent}
  j^\mu = {\nabla}_\sigma \left( {\nabla}^\mu \xi^\sigma - {\nabla}^\sigma \xi^\mu \right) .
\end{equation}

\noindent Threfore, the Weiss variation of the Einstein-Hilbert action, as in \ref{WAC-ACFActionVar1}, may then be written in the form:
\begin{equation} \label{WAC-ACFActionVar2}
  \begin{aligned}
  \Delta S_{\rm EH} =&~\frac{1}{2 \kappa} \biggl\{\int_{U} d^4 x  \sqrt{-g}  \, G_{\mu \nu}  \,\delta g^{\mu \nu} \\
  & \qquad + \int_{\partial U} d\Sigma_\mu
  \biggl[
    \delta{^{\mu \alpha}}{_{\nu \sigma}} \, g^{\sigma \beta} \, \Delta {\Gamma}{^\nu}{_{\alpha \beta}}
    - \left[2G{^\mu}{_\sigma} \, \xi^\sigma  - j^\mu \right] \Delta \lambda
  \biggr] \biggr\},
\end{aligned}
\end{equation}

\noindent where $G_{\mu \nu}$ is the Einstein tensor. Note that, the above equation suggests that the combination $\sqrt{-g}g^{\sigma \beta}$ to be the momentum conjugate to $\delta^{\mu \alpha}_{\,\,\,\,\,\,\,\nu \sigma}\delta \Gamma^{\nu}_{\alpha\beta}$ and hence is consistent with the findings of \cite{Parattu:2013gwa}. Additionally, the above result provides an interesting interpretation for $\left(2G^{\mu}_{\nu}\xi^{\nu}-j^{\mu}\right)\nabla_{\mu}\Phi$, as the generator for the translation in a direction perpendicular to the boundary surfaces, denoted by $\Phi=\textrm{constant}$ and is consistent with the Hamiltonian for general relativity, when $\Phi$ is chosen to be the time coordinate. 

\subsection{Boundary terms and their variation}

A conceptual difficulty with the Einstein-Hilbert action is that the elimination of the boundary terms requires boundary conditions in which both the metric and its normal derivatives must be held fixed, which is not in general consistent with the field equations. This difficulty is often dealt with by adding a suitable boundary term to the action, thereby canceling the normal derivatives of the metric \cite{Parattu2015}. We consider here a non-covariant boundary term of the following form:
\begin{equation} \label{WAC-BI1}
  \begin{aligned}
  B_1
  %
  &:= -\frac{1}{2 \kappa} \int_{U} d^4 x \, {\partial}_\mu\left\{\sqrt{-g}\left(\delta{^{\mu \sigma}}{_{\nu \alpha}} \, g^{\alpha \beta} \, {\Gamma}{^\nu}{_{\sigma \beta}}\right)\right\} \\
  &\,= -\frac{1}{2 \kappa} \int_{\partial U} d\underline{\Sigma}_\mu \left\{\sqrt{-g} \, \delta{^{\mu \sigma}}{_{\nu \alpha}} \, g^{\alpha \beta} \, {\Gamma}{^\nu}{_{\sigma \beta}}\right\} ,
  \end{aligned}
\end{equation}

\noindent which can be used to obtain the familiar ``${\Gamma} {\Gamma}$'' form of the action:
\begin{equation} \label{WAC-GravAction}
  \begin{aligned}
  S_{{\Gamma} {\Gamma}} &= S_{\rm EH} + B_1 \\
  &= \frac{1}{2 \kappa}\int_{U} d^4 x \sqrt{-g} \, g^{\alpha \beta} \biggl[ {\Gamma}{^\sigma}_{\beta \tau} \, {\Gamma}{^\tau}_{\sigma \alpha} - {\Gamma}{^\sigma}{_{\sigma \mu}}  {\Gamma}{^\mu}{_{\alpha \beta}} \biggr].
  \end{aligned}
\end{equation}

\noindent Neglecting boundary displacements, the variation takes the form:
\begin{equation} \label{WAC-NCBI1var}
  \begin{aligned}
  \delta B_1
  &= -\frac{1}{2 \kappa} \int_{\partial U}  d\underline{\Sigma}_\mu \, \sqrt{-g}
  \biggl\{g^{\tau \rho}\,\delta{^{\mu \sigma}}{_{\nu \tau}} \, \delta {\Gamma}{^\nu}{_{\sigma \rho}}
  \\
  &+
    \left[
    \delta{^{\mu \sigma}}{_{\nu \alpha}} \, {\Gamma}{^\nu}{_{\sigma \beta}}
    -
    \frac{1}{2}\left(\delta{^{\mu \sigma}}{_{\nu \tau}} g^{\tau \rho} \, {\Gamma}{^\nu}{_{\sigma \rho}}\right)
    \,
    g_{\alpha \beta}
    \right]
    \delta g^{\alpha \beta}
  \biggr\},
\end{aligned}
\end{equation}

\noindent which for vanishing boundary variation, i.e., for $\delta x^{\mu}=0$, cancels the term involving $\delta {\Gamma}{^\sigma}{_{\mu \nu}}$ in the variation of $\delta S_{\rm EH}$, presented in \ref{WAC-ACFActionVar1}. This leads to a well-posed boundary value problem for $S_{\Gamma \Gamma}$ action\footnote{The connection between $B_{1}$, the GHY boundary term and the null boundary term can be found in \cite{Parattu:2016trq}.}.

The cost, of course, is that the boundary term $B_1$ is non-covariant. A standard way around this is to construct a geometric boundary term that coincides with $B_1$ in an appropriate gauge; for spacelike and timelike boundary surfaces, this is provided by the Gibbons-Hawking-York (GHY) boundary term, and generalizations have been formulated for boundary surfaces which are null or have corners \cite{Parattu2015,Jubb2016,Parattu:2016trq}. Here, we consider an alternative approach, which maintains covariance and is valid for boundaries of arbitrary signature. This is achieved by introducing an additional boundary term $B_2$ dependent on the auxiliary connection $\bar{\Gamma}{^\lambda}{_{\mu \nu}}$:
\begin{equation} \label{WAC-BI2}
  B_2 := \frac{1}{2 \kappa}\int_{U} d^4 x \, {\partial}_\mu\left\{\sqrt{-g}\left(\delta{^{\mu \sigma}}{_{\nu \alpha}} \, g^{\alpha \beta} \, \bar{\Gamma}{^\nu}{_{\sigma \beta}}\right)\right\} .
\end{equation}

\noindent This approach is equivalent to that of the connection independent form of the gravitational action \cite{Harada2020CIGA}, and also closely related to the ``general perturbation" formalism \cite{Katzetal1997,Lynden-Belletal1995}.

To preserve general covariance, one then constructs a boundary term of the following form:
\begin{equation} \label{WAC-BI}
  \begin{aligned}
    B & := B_1 + B_2 \\
      &\,= -\frac{1}{2 \kappa} \int_{\partial U} d\underline{\Sigma}_\mu \left(\sqrt{-g} \, W^\mu\right),
  \end{aligned}
\end{equation}

\noindent where the following quantities are defined:
\begin{equation} \label{WAC-Wvect}
  W^\mu := \delta{^{\mu \alpha}}{_{\nu \sigma}} \, g^{\sigma \beta} \, W{^\nu}{_{\alpha \beta}}
\end{equation}
\begin{equation}\label{WAC-WDiff}
  W{^\lambda}{_{\mu \nu}} := {\Gamma}{^\lambda}{_{\mu \nu}} - \bar{\Gamma}{^\lambda}{_{\mu \nu}}.
\end{equation}

\noindent The quantity $W{^\lambda}{_{\mu \nu}}$ is the contorsion tensor, from which one can construct two scalars:
\begin{equation}\label{WAC-RicciScalarHarada}
\bar{R} := g^{\mu \nu} \bar{R}_{\mu \nu}
\end{equation}
\begin{equation}\label{WAC-WScalar}
W^2 := W{^{\mu \nu}}{_\lambda} W{^\lambda}{_{\mu \nu}} - W{^{\nu \mu}}{_\mu} W{^\lambda}{_{\lambda \nu}} ,
\end{equation}

\noindent where $\bar{R}_{\mu \nu}$ is the Ricci tensor constructed from $\bar{\Gamma}{^\lambda}{_{\mu \nu}}$. At this stage, it is worth mentioning the properties of the contorsion tensor $W{^\lambda}{_{\mu \nu}}$ and its relation with the anti-symmetric part of the connection. If both $\Gamma{^\mu}{_{\alpha \beta}}$ and $\bar{\Gamma}{^\mu}{_{\alpha \beta}}$ arise from metric tensors alone, i.e., if both of these are affine connections, then $W{^\mu}{_{\alpha \beta}}$ would be non-zero and symmetric in the lower indices. Thus, even in the absence of any torsion, the tensor $W{^\mu}{_{\alpha \beta}}$ would be non-zero. However, in the presence of torsion, the contorsion tensor $W{^\lambda}{_{\mu \nu}}$ will not be symmetric, since if $\bar{\Gamma}^{\mu}_{\alpha \beta}$ includes torsional degrees of freedom, then $W{^\mu}{_{[\alpha \beta]}}$ will coincide with the torsion tensor. As we will observe, this contorsion tensor will play the key role in the subsequent analysis.

\subsection{Modified gravitational action and its Weiss variation}

We have described, in the earlier sections, the Weiss variation of the Einstein-Hilbert action and have provided a covariant boundary term necessary to make the variational problem well-posed. Here we construct a covariant gravitational action $S_{\rm gW}$, using the covariant boundary term $B$, such that
\begin{equation} \label{WAC-ACFAction4}
  S_{\rm gW} = S_{\rm EH} + B ,
\end{equation}

\noindent which can be shown to be equivalent to the following forms \cite{Lynden-Belletal1995,Harada2020CIGA}:
\begin{equation} \label{WAC-ACFActionEquiv}
\begin{aligned}
S_{\rm gW}
    &= \frac{1}{2 \kappa}\int_{U} d^4 x  \sqrt{-g} \left[{R} - {\nabla}_\mu W^\mu\right]
    \\
    &= \frac{1}{2 \kappa}\int_{U} d^4 x  \sqrt{-g} \left[{R} - {\nabla}_\mu(W{^{\mu \nu}}{_\nu}-W{_\nu}{^{\nu \mu}})\right] 
    \\
    &= \frac{1}{2 \kappa}\int_{U} d^4 x  \sqrt{-g} \left(\bar{R}+W^2\right).
\end{aligned}
\end{equation}

\noindent where $\bar{R}$ has been defined in \ref{WAC-RicciScalarHarada}. One may interpret the non-dynamical part of the boundary term $B$, in particular $B_2$ (see \ref{WAC-BI2} for an explicit expression), to be a subtraction term for the gravitational action. Since the connection $\bar{\Gamma}^{\mu}_{\alpha \beta}$ is general, $B_2$ can be constructed with respect to a rather general class of reference geometries (which need not even correspond to Riemannian geometries), and does not require that the boundary surface be isometrically embeddable in some reference spacetime (as opposed to the case of the GHY subtraction term). For instance, in the case of an asymptotically flat spacetime, one can simply use for $\bar{\Gamma}{^\gamma}{_{\mu \nu}}$ the Christoffel symbols for flat spacetime in an appropriate set of coordinates for the boundary surfaces at infinity. More generally, one requires that the reference spacetime has the same asymptotic and topological structure as the spacetime domain of interest.\footnote{It is worth noting that a similar approach was originally proposed in \cite{IsenbergNester1980} (in that case the subtraction term was constructed from a flat connection) and employed in a simplified proof of the positive energy theorem \cite{Nester1982}.}

To better understand this point, first note that when evaluating on the same connection $B_1$ and $B_2$ differ by a sign. Now consider a manifold $M = w_1 \cup w_2$ which is compact and without boundary, such that the regions $w_1$ and $w_2$ share a common boundary $\partial w$. Since $M$ is without boundary, one has (using \ref{WAC-GravAction}):
\begin{equation} \label{WAC-GravActionM}
\begin{aligned}
S_{{\rm EH}(M)} &= S_{\Gamma \Gamma (M)} = S_{{\Gamma} {\Gamma}(w_1)} + S_{{\Gamma} {\Gamma}(w_2)} \\
& = S_{{\rm EH}(w_1)} + B_{1(\partial w_1)} + S_{{\rm EH}(w_2)} + B_{1(\partial w_2)}.
\end{aligned}
\end{equation}

\noindent where $S_{{\Gamma} {\Gamma}(w_1)}$ is $S_{{\Gamma} {\Gamma}}$ evaluated on $w_1$ and $S_{{\Gamma} {\Gamma}(w_2)}$ is $S_{{\Gamma} {\Gamma}}$ evaluated on $w_2$. Both $S_{{\Gamma} {\Gamma}(w_1)}$ and $S_{{\Gamma} {\Gamma}(w_2)}$ come with the respective boundary terms $B_{1(\partial w_1)}$ and $B_{1(\partial w_2)}$, with opposite orientations. For smooth metrics, these boundary terms cancel, though they may differ for $C^0$ metrics (for instance, if the coordinate charts are $C^0$ at $\partial w$, or if one considers geometries that are ``glued'' at $\partial w$). If one is primarily concerned with the variations in the region $w_1$, then one may combine the boundary terms $B_{1(\partial w_1)}$ and $B_{1(\partial w_2)}$ to form a covariant boundary term for the action on $w_2$ which is nontrivial if the metric is only $C^0$ at the boundary. Upon comparison with \ref{WAC-BI}, one then recovers the interpretation of the boundary term $B_2$ in \ref{WAC-BI} as a subtraction term associated with an external reference geometry.

Another way to view this procedure (in particular the addition of boundary terms) is as a generalization of the canonical transformation in mechanics. In the mechanical case, one constructs a generating functional with the appropriate arguments, and then adds a total (time) derivative of the generating functional to the action. The additional endpoint terms will correspond to a different identification of the conjugate variables in the resulting Weiss variation; the difference is in fact a canonical transformation for the appropriate choice of generating functions. The addition of the boundary term $B_1$ has a similar effect (a similar observation was made in \cite{BrownLauYork2000,BrownYork1992}), and corresponds to the viewpoint that the (inverse) metric variables contain the degrees of freedom for general relativity.

The full variation of $S_{\rm EH}$ (including boundary displacements) was obtained in \ref{WAC-ACFActionVar2}. The full Weiss variation of the boundary term $B$ may be obtained (see \ref{WAC-BdyIntVar4} in \ref{Appcovdiv} for details), which takes the form:
\begin{equation} \label{WAC-BdyCovIntVar}
  \begin{aligned}
    \Delta B
    &= -\int_{\partial U} \frac{d\underline{\Sigma}_\sigma}{2 \kappa} \, \sqrt{-g} \left[ \Delta W^\sigma - \frac{1}{2} \, W^\sigma \, g_{\mu \nu} \, \Delta g^{\mu \nu} + \hat{j}^\mu \right] \\
    &= -\int_{\partial U} \frac{d{\Sigma}_\sigma}{2 \kappa}
    \biggl[
      \delta{^{\sigma \alpha}}{_{\rho \tau}} g^{\tau \beta} \Delta W{^\rho}{_{\alpha \beta}} - P{^\sigma}_{\mu \nu} \Delta g^{\mu \nu} + \hat{j}^\mu
    \biggr]
    .
  \end{aligned}
\end{equation}

\noindent where the current $\hat{j}^{\mu}$ and the momentum $P{^\sigma}_{\mu \nu}$, symmetric in the lower two indices are defined as:
\begin{equation} \label{WAC-JdefII}
  \hat{j}^{\mu} := \nabla_\sigma
  \left(
  \xi^\mu \, W{^\sigma} - \xi^\sigma \, W{^\mu}
  \right),
  \end{equation}
\begin{equation} \label{WAC-Momentum}
  P{^\sigma}_{\mu \nu} := - \delta{^{\sigma \alpha}}{_{\rho (\mu}} \, W{^\rho}{_{\alpha |\nu)}} + \frac{1}{2} \, W^\sigma \, g_{\mu \nu} .
\end{equation}

\noindent Using the above variation of the boundary term $B$, one may obtain the full Weiss variation of the gravitational action $S_{\rm gW}$ as:
\begin{equation} \label{WAC-ACFActionVarFull}
  \begin{aligned}
  \Delta S_{\rm gW} =&~\frac{1}{2 \kappa}
  \biggl\{
    \int_{U} d^4 x  \sqrt{-g} \, G_{\mu \nu} \, \delta g^{\mu \nu} + \int_{\partial U} d\Sigma_\mu \biggl[ {\Xi}^\mu 
    +
    P{^\mu}_{\sigma \beta} \, \Delta g^{\sigma \beta}
    -
    \left[2G{^\mu}{_\sigma}\, \xi^\sigma - J^\mu \right] \Delta \lambda \, \biggr]
  \biggr\},
  \end{aligned}
\end{equation}

\noindent where $J^\mu := j^\mu + \hat{j}^\mu$, and the following quantity has been defined:
\begin{equation} \label{WAC-XiDef}
  \begin{aligned}
    {\Xi}^\mu &:=
      \delta{^{\mu \alpha}}{_{\nu \sigma}} \, g^{\sigma \beta} \,
      \Delta \bar{\Gamma}{^\nu}{_{\alpha \beta}} .
  \end{aligned}
\end{equation}

We will now discuss a useful geometrical identity, sometimes referred to in the literature as the Katz-Ori identity \cite{Lynden-Belletal1995}, which establishes an expression for $J^\mu = j^\mu + \hat{j}^\mu$. For the purpose of deriving the desired identity, we start by computing the Lie derivative of $W{^\alpha}{_{\mu \nu}}$, which takes the form:
\begin{equation} \label{WAC-LDWdef}
\begin{aligned}
\pounds_\xi W^\alpha{_{\mu \nu}}
=&~ \pounds_\xi {\Gamma}^\alpha{_{\mu \nu}} - \pounds_\xi \bar{\Gamma}^\alpha{_{\mu \nu}} \\
=&~ {\nabla}_\mu {\nabla}_\nu \xi^\gamma + \xi^\sigma {R}{^\gamma}_{\nu \sigma \mu} - \pounds_\xi \bar{\Gamma}^\alpha{_{\mu \nu}} .
\end{aligned}
\end{equation}

\noindent Defining the following geometrical quantity, $\tilde{\Xi}^\mu := \Delta \lambda^{-1} ({\Xi}^\mu - \delta\bar{\Gamma}^\mu)$, where, $\delta\bar{\Gamma}^\mu := \delta{^{\mu \alpha}}{_{\nu \sigma}} \delta \bar{\Gamma}^\nu{_{\alpha \beta}}$, it follows that,
\begin{equation} \label{WAC-XiLD}
  \begin{aligned}
  \tilde{\Xi}^\mu
  =&~
  -
  \delta{^{\mu \alpha}}{_{\nu \sigma}} g^{\sigma \beta} \pounds_\xi W^\nu{_{\alpha \beta}} + 2 {R}{^\mu}{_\sigma} \xi^\sigma - j^\mu.
  \end{aligned}
\end{equation}

\noindent Further manipulations involving the Lie derivative of the contorsion tensor yields,
\begin{equation} \label{WAC-LDWdefII}
  \delta{^{\mu \alpha}}{_{\nu \sigma}} g^{\sigma \beta} \pounds_\xi W^\nu{_{\alpha \beta}}
  = \hat{j}^{\mu} - \Sigma^{\mu \alpha}{_\beta} {\nabla}_\alpha \xi^\beta - \xi^\mu {\nabla}_\sigma W^\sigma,
\end{equation}

\noindent where:
\begin{equation} \label{WAC-udef}
\begin{aligned}
\Sigma^{\mu \alpha}{_\beta} := ~&
g^{\mu \beta} \, W{^\sigma}_{\sigma \beta} + \delta{_\beta}{^\mu} \, W{^\sigma}{_\sigma}{^\alpha} - W{^\mu}{_\beta}{^\alpha} \\
& - W^{\mu \alpha}{_\beta} + \left( W^{\mu \sigma}{_\sigma} - W{^\sigma}{_\sigma}{^\mu} \right) \delta{^\alpha}{_\beta}.
\end{aligned}
\end{equation}

\noindent Using the above expression for the Lie derivative of the contorsion tensor, one may show that \ref{WAC-XiLD} may be rewritten in the following form:
\begin{equation} \label{WAC-XiLD2}
  \begin{aligned}
    J^{\mu} = j^\mu + \hat{j}^\mu = 
    \Sigma^{\mu \alpha}{_\beta} {\nabla}_\alpha \xi^\beta + \xi^\mu {\nabla}_\sigma W^\sigma + 2 {R}{^\mu}{_\sigma} \xi^\sigma - \tilde{\Xi}^\mu .
  \end{aligned}
\end{equation}

\noindent With some work, one may show that the term $\Sigma^{\mu \alpha}{_\beta} {\nabla}_\alpha \xi^\beta$ can be reexpressed as:
\begin{equation} \label{WAC-SpinTerm}
  \begin{aligned}
  \Sigma^{\mu \alpha}{_\beta} {\nabla}_\alpha \xi^\beta
  =&~ \Sigma^{\mu \alpha}{_\beta} \bar{\nabla}_\alpha \xi^\beta  
  \\
  &+ \xi^\beta \left[ - {_{\rm E}}\Theta{^\mu}{_\beta} + \sigma{^\mu}{_\beta} - \delta{^\mu}{_\beta} L_{\rm gW} \right],
  \end{aligned}
\end{equation}

\noindent where $L_{\rm gW} := {R} - {\nabla}_\sigma W^\sigma$ and ${_{\rm E}}\Theta{^\alpha}{_\beta}$ is a generalization of the Einstein energy-momentum pseudotensor\footnote{Such a quantity was previously obtained in \cite{Rosen1940} for a flat reference connection and in \cite{Lynden-Belletal1995} for a general metric compatible reference connection.}:
\begin{equation} \label{WAC-EinsteinEMPPT}
\begin{aligned}
{_{\rm E}}\Theta{^\alpha}{_\beta}=
~
&
W{^\sigma}_{\beta \sigma}  W{_\tau}^{\tau \alpha}
-
W^{\alpha \tau}{_\tau} W{^\sigma}_{\beta \sigma}
\\
&
-
W{^\sigma}_{\sigma \tau} \left( W{^\alpha}{_\beta}{^\tau} + W{^\tau}{_\beta}{^\alpha}
\right)\\
&
+
W^{\alpha \sigma \tau} \left( W_{\sigma \beta \tau} + W_{\tau \beta \sigma} \right)
-
\delta{^\alpha}{_\beta} \,L_{\rm gW}
,
\end{aligned}
\end{equation}

\noindent and $\sigma{_\beta}{^\alpha}$ is a quantity constructed from the torsion tensor $\bar{T}{^\mu}_{\alpha \beta}$, and the tensor $C_{\mu \alpha \beta} = \bar{\nabla}_\mu g_{\alpha \beta}$:
\begin{equation} \label{WAC-SigTors}
\begin{aligned}
\sigma{^\alpha}{_\beta}=~
&
\frac{1}{2} C^{\sigma \tau}{_\tau} \left[ \bar{T}{^\alpha}_{\beta \sigma}
-
\bar{T}{_\sigma}{^\alpha}{_\beta} \right]
+
\bar{T}_{\sigma \beta \tau} \left[ C^{\alpha \sigma \tau} + \bar{T}^{\sigma \alpha \tau} \right]\\
&
-
C^{\sigma \alpha \tau} \left[ \bar{T}_{\sigma \beta \tau} + \bar{T}_{\tau \beta \sigma} \right]
+
\left[ C^{\sigma \alpha}{_\sigma} - C^{\alpha \sigma}{_\sigma} \right] \bar{T}{^\tau}_{\beta \tau} \\
&
+
\left[ \bar{T}{^\alpha}{_\beta}{^\sigma} - \bar{T}^{\sigma \alpha}{_\beta} \right] \bar{T}{^\tau}_{\sigma \tau}
+
\bar{T}^{\sigma \alpha \tau} T_{\tau \beta \sigma} \\
&
- 2 \bar{T}^{\sigma \alpha}{_\sigma} \bar{T}{^\tau}_{\beta \tau} .
\end{aligned}
\end{equation}

\noindent Observe that when the torsion tensor vanishes, the quantity $\sigma{_\beta}{^\alpha}$ does not contribute. Inserting the above into \ref{WAC-XiLD2}, we obtain the Katz-Ori identity:
\begin{equation} \label{WAC-KOri}
  \begin{aligned}
  J^\mu
  = \, &
  \xi^\beta \left[2 G{^\mu}{_\beta} - {_{\rm E}}\Theta{^\mu}{_\beta} + \sigma{^\mu}{_\beta} \right]
  +\Sigma^{\mu \alpha}{_\beta} \bar{\nabla}_\alpha \xi^\beta - \tilde{\Xi}^\mu ,
  \end{aligned}
\end{equation}

\noindent which may be rewritten as:
\begin{equation} \label{WAC-XiLD4}
  \begin{aligned}
  2 {G}{^\mu}{_\beta} \xi^\beta - \tilde{\Xi}^\mu - J^\mu
  = \, &
  \xi^\beta \left[ {_{\rm E}}\Theta{^\mu}{_\beta} - \sigma{^\mu}{_\beta} \right]
  - \Sigma^{\mu \alpha}{_\beta} \bar{\nabla}_\alpha \xi^\beta.
  \end{aligned}
\end{equation}

\noindent With this result in hand, one may then obtain the full Weiss variation of the action $S_{\rm gW}$:
\begin{equation} \label{WAC-ACFActionVarFullXi}
  \begin{aligned}
  \Delta S_{\rm gW} =&~\frac{1}{2 \kappa}
  \biggl\{
    \int_{U} d^4 x  \sqrt{-g} G_{\mu \nu}\delta g^{\mu \nu}
    \\
    &+ \int_{\partial U} d\Sigma_\mu \biggl[
    P{^\mu}_{\sigma \beta} \Delta g^{\sigma \beta}
    + \delta\bar{\Gamma}^\mu
    \\
    &-
    \left[( {_{\rm E}}\Theta{^\mu}{_\beta} - \sigma{^\mu}{_\beta} )\xi^\beta
    - \Sigma^{\mu \alpha}{_\beta} \bar{\nabla}_\alpha \xi^\beta \right] \Delta \lambda
    \biggr]
  \biggr\}.
  \end{aligned}
\end{equation}

\noindent If $\bar{\nabla}_\alpha$ is torsion-free, then $\sigma{^\mu}{_\nu}=0$. In addition one may choose the reference connection $\bar{\nabla}_\alpha$ to be the flat connection, with $\xi^\mu$ being a covariantly constant vector.\footnote{Alternatively, if one chooses $\xi^\mu$ to be a Killing vector with respect to a reference geometry, then for $\chi_\mu:=\eta_{\mu \nu} \xi^\nu$, one may write $\bar{\nabla}_{(\mu} \xi_{\nu)}=0$. It follows that $\bar{\nabla}_\alpha \xi^\beta = g^{\tau \beta} (\bar{\nabla}_\alpha \chi_\tau - C_{\alpha \sigma \tau} \,\xi^\sigma)$, so that (making use of $\Sigma^{\mu \alpha \beta} = \Sigma^{\mu \beta \alpha}$), one may write $\Sigma^{\mu \alpha}{_\beta} \, \bar{\nabla}_\alpha \xi^\beta = - \Sigma^{\mu \alpha \sigma} C_{\alpha \beta \sigma} \,\xi^\beta$.} One may also require that the variation of the auxiliary connection vanishes so that $\delta\bar{\Gamma}^\mu=0$ (this is equivalent to the statement that $\Delta \bar{\Gamma}{^\nu}{_{\alpha \beta}}$ is induced purely by boundary displacements). Under these assumptions, one obtains the simplified Weiss variation of the gravitational action:
\begin{equation} \label{WAC-ACFActionVarFullRW5}
  \begin{aligned}
  \Delta S_{gW} =&\frac{1}{2 \kappa} \biggl\{\int_{U} d^4 x  \sqrt{-g} G_{\mu \nu} \delta g^{\mu \nu} + \int_{\partial U} d\Sigma_\mu \biggl[ P{^\mu}_{\alpha \beta} \Delta g^{\alpha \beta}
  -{_{\rm E}}\Theta{^\mu}{_\beta} \, \xi^\beta\,\Delta \lambda
  \biggr]
  \biggr\}.
  \end{aligned}
\end{equation}

\noindent Upon comparison with the field theoretical case, as in \ref{WAC-WeissVariationFields}, we see that the tensor ${_{\rm E}}\Theta{^\mu}{_\beta}$ serves as a generalization of the canonical energy-momentum tensor for the gravitational case. This being the reason to refer to ${_{\rm E}}\Theta{^\mu}{_\beta}$ as the generalization of the Einstein energy-momentum pseudotensor. 

Also note that the expression for the Noether current is independent of the variational principle being considered, but depends on the structure of the Lagrangian and, in particular, on the boundary terms. To see this, first consider the normal variation of the Einstein-Hilbert action, such that the boundary contribution due to \ref{WAC-ACFActionVar2} becomes, $\delta{^{\mu \alpha}}{_{\nu \sigma}} \, g^{\sigma \beta} \, \delta {\Gamma}{^\nu}{_{\alpha \beta}}$. If we now consider the variation to be due to ``passive" diffeomorphism, such that the coordinates are changed alone, but not the boundary, then the on-shell variation of the action will be given by the boundary term dependent on $j^{\mu}$ alone, as evident from \ref{WAC-LieDerivConnectioncont}. Consider now the case of Weiss variation of the Einstein-Hilbert action, which may be considered as an ``active" diffeomorphism, as the boundary also gets modified. Choosing $\Delta \Gamma{^\nu}{_{\alpha \beta}}=0$ in the boundary term --- this may be interpreted as the statement that the field values $\Gamma{^\nu}{_{\alpha \beta}}$ are Lie transported with the boundary --- one obtains the on-shell (in this case $G_{\mu \nu}=0$) boundary variation (in this case the coefficient of $\Delta \lambda$) to be given by $j^{\mu}$ alone. In which case the Noether current is given by $j^{\mu}$, which has an identical expression to the one derived from the diffeomorphism invariance of general relativity. On the other hand, the Weiss variation of the gravitational action $S_{\rm gW}$, presented in \ref{WAC-ACFActionVarFull}, defines a different Noether current $J^{\mu}$, depending on the contorsion tensor $W{^{\mu}}{_{\alpha \beta}}$. The reason for the difference between $J^{\mu}$ and $j^{\mu}$ is due to the fact that the Einstein-Hilbert action and $S_{\rm gW}$ differ by the boundary term $B$, whose Weiss variation depends precisely on $(J^{\mu}-j^{\mu})$ (see \ref{WAC-BdyCovIntVar} for further details). Thus the structure of the Noether current is more sensitive to the structure of the gravitational Lagrangian, than the nature of the variational principle itself.


%
%

%
%
\section{Geometric interpretation of boundary terms}\label{geometry_boundary}

In this section, we show that the momenta conjugate to the metric $g^{\mu \nu}$ correspond to well-known geometric expressions in the appropriate gauges. Since the reference connection $\bar{\Gamma}{^\sigma}_{\mu \nu}$ is non-dynamical, one may for simplicity choose it to vanish without affecting the dynamics; we make this choice to simplify the computations. However, we will demonstrate that the reference connection is indeed linear in the conjugate momenta, so one can readily recover the full expressions with non-zero $\bar{\Gamma}{^{\mu}}{_{\alpha \beta}}$. With regard to the Hamiltonian, one can recover in a straightforward fashion the earlier result of \cite{FengMatzner2018}; from \ref{WAC-ACFActionVarFull}, we see that if one disregards the total derivative term $J^\mu$ (which is appropriate for compact boundary surfaces), the Hamiltonian for vacuum GR becomes a combination of Gauss-Codazzi constraints arising from the projection of the Einstein tensor $G{^\mu}_\nu$ onto the boundary surfaces. It is worth noting that \ref{WAC-ACFActionVarFull} holds for null surfaces as well, and one finds via a similar argument that the term proportional to $\Delta \lambda$ in the null case becomes a combination of the analogous constraints for null hypersurfaces \cite{Brady1995,*Torre1986,*DInverno1980}. In what follows, we work out in detail the expressions for the conjugate momenta.

\subsection{General expression}

In general, the boundary element $d\Sigma_\mu$ is proportional to some vector $v^\mu$, so that the conjugate momentum takes the form (note that the symmetry in the lower indices of $P{^\sigma}_{\mu \nu}$ is induced by the symmetry in $\Delta g^{\mu \nu}$):
\begin{equation} \label{WAC-ConjMom0}
  \begin{aligned}
  P_{\mu \nu} = v_\sigma \, P{^\sigma}_{\mu \nu} = - v_\sigma \, \delta{^{\sigma \alpha}}{_{\rho (\mu}} \, W{^\rho}{_{\alpha |\nu)}} + \frac{1}{2} \, v_\sigma \, W^\sigma \, g_{\mu \nu} .
  \end{aligned}
\end{equation}

\noindent Using the result that $W{^{\mu}}{_{\alpha \beta}}=\Gamma{^{\mu}}{_{\alpha \beta}}-\bar{\Gamma}{^{\mu}}{_{\alpha \beta}}$, and the fact that the above expression for the conjugate momentum is linear in $W{^{\mu}}{_{\alpha \beta}}$, it follows that the conjugate momentum will be linear in both $\Gamma{^{\mu}}{_{\alpha \beta}}$ and $\bar{\Gamma}{^{\mu}}{_{\alpha \beta}}$, individually. Thus we can set the reference connection to be zero, which be easily restored in the expression for the conjugate momentum. Considering the reference connection to be vanishing, the above expression for the conjugate momentum can be obtained:
\begin{equation} \label{WAC-ConjMom1}
\begin{aligned}
P_{\mu \nu} = \,
& v_\sigma \, \Gamma{^\sigma}_{\mu \nu} 
- v_{(\mu} \Gamma{^\sigma}_{\sigma | \nu)}
+ \frac{1}{2} g_{\mu \nu} v^\tau \left[\Gamma{^\sigma}_{\sigma \tau} - \Gamma_{\tau \sigma}{^\sigma}\right].
\end{aligned}
\end{equation}

\noindent One may make use of the following expressions:
\begin{equation} \label{WAC-DetGradientandFriends}
  \begin{aligned}
  {\Gamma}{^\sigma}_{\sigma \mu} &= \partial_\mu \ln(\sqrt{-g}) \\
  \Gamma_{\mu \sigma}{^\sigma} &= -\partial_\mu \ln\sqrt{-g} + g^{\alpha \beta} \partial_\alpha g_{\mu \beta}
\end{aligned}
\end{equation}

\noindent to obtain:
\begin{equation} \label{WAC-ConjMom2}
  \begin{aligned}
  P_{\mu \nu} = \,
  & v_\sigma \, \Gamma{^\sigma}_{\mu \nu} 
  - v_{(\mu} \partial_{\nu)} (\ln\sqrt{-g}) \\
  &
  + \frac{1}{2} g_{\mu \nu} v^\tau \left[2\partial_{\tau} (\ln\sqrt{-g}) - g^{\alpha \beta} \partial_\alpha g_{\tau \beta}\right].
\end{aligned}
\end{equation}

\noindent Owing to the fact that the conjugate momentum is linear in $W{^\lambda}{_{\mu \nu}}$, one can readily recover the contribution of a non-vanishing reference connection $\bar\Gamma{^\sigma}_{\mu \nu}$ by adding the following terms to \ref{WAC-ConjMom1}:
\begin{equation} \label{WAC-MomRefConn}
  \begin{aligned}
    v_{(\mu} \bar\Gamma{^\sigma}_{\sigma | \nu)}
    - v_\sigma \, \bar\Gamma{^\sigma}_{\mu \nu} 
    - \frac{1}{2} g_{\mu \nu} v^\tau \left[\bar\Gamma{^\sigma}_{\sigma \tau} - \bar\Gamma_{\tau \sigma}{^\sigma}\right].
  \end{aligned}
\end{equation}

\noindent In order to proceed any further, we need to specify the vector field $v_{\sigma}$, which requires fixing the nature of the boundary surface. In what follows we first consider the case of spacelike/timelike surfaces and then discuss the case of null surfaces. In particular, we will demonstrate that in all of these cases the conjugate momentum $P_{\mu \nu}$ has an interesting geometrical interpretation.

\subsection{Timelike and spacelike boundaries}

Here, we consider the case where the boundaries are either strictly timelike or spacelike. In this case, the boundary element $d\Sigma_\mu$ is proportional to a timelike or spacelike normalized vector field $n_\mu$. In terms of the lapse $\alpha$ and the shift vector $\beta^i$, the unit normal vector $n^\mu$ has the components:
\begin{equation} \label{WAC-UnitNormal}
\begin{aligned}
n^\cdot &= \alpha^{-1} \left(1,\beta^\cdot \right)  \\
n_\cdot &= \left(\varepsilon \, \alpha, 0, 0, 0\right) ,
\end{aligned}
\end{equation}

\noindent where the dot indicates index placement, and $\varepsilon = \pm 1$ indicates the sign of the norm for $n$. The extrinsic and mean curvatures can be computed according to the formula (the acceleration is given by $a_\nu = -\varepsilon D_\nu \ln \alpha$):
\begin{equation} \label{WAC-Extrinsic}
\begin{aligned}
K_{\mu \nu} &= {\nabla}_\mu n_\nu - \varepsilon \, n_\mu \, a_\nu = \partial_\mu n_\nu - n_\sigma \, {\Gamma}{^\sigma}_{\mu \nu} - \varepsilon \, n_\mu \, a_\nu \\
K &= {\nabla}_\sigma n^\sigma = \partial_\sigma n^\sigma + n^\nu \, {\Gamma}{^\sigma}_{\sigma \nu}.
\end{aligned}
\end{equation}

\noindent On timelike and spacelike boundaries, one may choose Gaussian normal coordinates, in which the lapse $\alpha$ and the shift vector $\beta^i$ are constant and can be taken to have the values $\alpha=1$ and $\beta^i=0$. One has the following relations between the extrinsic curvature and the connection:
\begin{equation} \label{WAC-ExtrinsicGaussian}
\begin{aligned}
K_{\mu \nu} &= - n_\sigma \, {\Gamma}{^\sigma}_{(\mu \nu)} \\
K &=  n^\tau \, {\Gamma}{^\sigma}_{\sigma \nu}  
= -  n^\tau \, {\Gamma}_{\tau \sigma}{^\sigma} \, .
\end{aligned}
\end{equation}

\noindent Plugging the above into \ref{WAC-ConjMom1}, the conjugate momentum takes the form (making use of $\sqrt{-g}=\alpha \sqrt{\gamma}$):
\begin{equation} \label{WAC-ConjMomsl}
\begin{aligned}
P_{\mu \nu} = K_{\mu \nu} - \varepsilon \, K \, n_\nu \, n_\mu - K \,  \gamma_{\mu \nu},
\end{aligned}
\end{equation}

\noindent where $\gamma_{\mu \nu} = g_{\mu \nu} - \varepsilon n_\mu n_\nu$ is the projection operator. If one restricts to variations of the inverse induced metric on the boundary, then:
\begin{equation} \label{WAC-ConjMomPdq}
\begin{aligned}
P_{\mu \nu} \Delta g^{\mu \nu} = p_{ij} \Delta \gamma^{ij},
\end{aligned}
\end{equation}

\noindent where $p_{ij}$ is the well-known 3+1 expression for the conjugate momentum:
\begin{equation} \label{WAC-ConjMomPsp}
\begin{aligned}
p_{ij} = K_{ij} - K \, \gamma_{ij} .
\end{aligned}
\end{equation}

\noindent The quantity $P_{\mu \nu}$ may then be interpreted as a generalization of the conjugate momentum to the inverse metric $\gamma^{ij}$ in the 3+1 formalism. Thus the conjugate momentum is really the momentum conjugate to the dynamical degrees of freedom of general relativity, as far as the spacelike/timelike boundaries are considered. We will now demonstrate that this is the case for null boundaries as well.

\subsection{Null boundaries}

For null boundaries, one may decompose the inverse metric in the following manner:
\begin{equation} \label{WAC-InvMetNullDecomp}
g^{\mu \nu} = q^{\mu \nu} - l^\mu k^\nu - l^\nu k^\mu,
\end{equation}

\noindent where $l^\mu$ is the null normal vector tangent to the null boundary, and $k^\mu$ is an auxiliary null vector satisfying the normalization $l_\mu k^\mu = -1$, with $l^\mu$ and $k^\mu$ being transverse to $q^{\mu \nu}$, which is a projection operator which projects onto a two-surface. In this case, $d\Sigma_\mu$ is proportional to $l_\mu$, and the conjugate momentum takes the form:
\begin{equation} \label{WAC-ConjMomNull0}
P_{\mu \nu} = l_\sigma \, P{^\sigma}_{\mu \nu}.
\end{equation}

\noindent We now consider the quantity $P_{\mu \nu} \Delta g^{\mu \nu}$ appearing in the boundary term, which takes the following form:
\begin{equation} \label{WAC-Pdg0}
  \begin{aligned}
  P_{\mu \nu} \Delta g^{\mu \nu} & =  P_{\mu \nu} \Delta q^{\mu \nu} - 2 P_{\mu \nu} l^\mu \Delta k^\nu  - 2 P_{\mu \nu} k^\mu \Delta l^\nu.
  \end{aligned}
\end{equation}

\noindent It is convenient to work in Gaussian null coordinates $\left(u,r,x,y\right)$, following the conventions of \cite{Parattu2015}, in which the line element takes the form:
\begin{equation} \label{WAC-LineElement}
  ds^2 = - 2r \alpha du^2 + 2 dr du - 2 r \beta_A du dx^A + q_{AB} dx^A dx^B ,
\end{equation}

\noindent where $r=0$ indicates the location of the boundary surface. Associated with this $r=0$ null surface, one can construct the null normal $\ell^{\mu}$ and the auxiliary null vector $k^{\mu}$ in the Gaussian null coordinates, which have the following expressions on the null surface: 
\begin{equation} \label{WAC-NullNormal}
  \begin{aligned}
  l^\mu &= \left(1,0,0,0\right)  \\
  k^\mu &= \left(0,-1,0,0\right).
\end{aligned}
\end{equation}

\noindent Using the above expression for the components of the null vectors on the null surface, one can show that $2 l_\sigma \Gamma{^\sigma}_{A B} =-\partial_u q_{AB}$, and $l^\tau g^{\alpha \beta} \partial_\alpha g_{\tau \beta} = -2 \alpha$ on the null boundary surface. With the volume element $\sqrt{-g}=\sqrt{q}$ for the above Gaussian null coordinates, one obtains from \ref{WAC-ConjMom2} the following expression:
\begin{equation} \label{WAC-Psurf}
  \begin{aligned}
  P_{A B} =\,  
  & - \theta_{AB} 
  + q_{AB} \left(\theta + \alpha \right),
  \end{aligned}
\end{equation}

\noindent where we have used the expressions for the extrinsic curvature $\theta_{AB}$, associated with the null normal of the null surface and its trace $\theta$:
\begin{equation} \label{WAC-NullExtrinsic}
  \begin{aligned}
    \theta_{AB} = \frac{1}{2} \partial_u q_{AB}
    \qquad \qquad
    \theta = \partial_u (\ln \sqrt{q}).
\end{aligned}
\end{equation}

\noindent One then has the expression:
\begin{equation} \label{WAC-PdgSurf}
  \begin{aligned}
  P_{\mu \nu} \Delta q^{\mu \nu} =
  P_{A B} \Delta q^{AB}
  = 
  \left[ \theta_{AB} 
  - q_{AB} \left(\theta + \alpha \right) \right]\Delta q^{AB}.
  \end{aligned}
\end{equation}

\noindent For the remaining terms, one can show that:
\begin{equation} \label{WAC-Pnorm}
  \begin{aligned}
  P_{\mu \nu} l^\mu =\,  
  & \frac{1}{2} \left[ \left(\theta + 2 \alpha\right) l_\mu - \nabla_\mu (l^2) \right] \\
  P_{\mu \nu} k^\mu =\,  
  & \left(\theta + \alpha\right) k_\mu - k^\nu \, \nabla_\mu l_\nu \\
  &+ \frac{1}{2} \left[ \partial_\mu\left(\ln\sqrt{-g}\right)-l_\mu k^\nu \partial_\nu \left(\ln\sqrt{-g}\right)\right],
  \end{aligned}
\end{equation}

\noindent and also:
\begin{equation} \label{WAC-Zeros}
  \begin{aligned}
  l_\mu \Delta l^\mu &= \Delta l^\mu \partial_\mu (\ln\sqrt{-g}) = l_\mu \Delta k^\mu = \Delta k^\mu \nabla_\mu (l^2) = 0 .
  \end{aligned}
\end{equation}

\noindent Combining all these expressions, the full conjugate momentum term given by \ref{WAC-Pdg0} from the Weiss variation of the gravitational action $S_{\rm gW}$, yields the following expression:
\begin{equation} \label{WAC-Pdg}
  \begin{aligned}
  P_{\mu \nu} \Delta g^{\mu \nu} & =  P_{AB} \Delta q^{AB} 
  \\
  &+ 2 \left[k^nu \nabla_\mu l_\nu - (\theta + \alpha) k_\mu\right]\Delta l^\mu,
  \end{aligned}
\end{equation}

\noindent where $P_{AB}\Delta q^{AB}$ is given by \ref{WAC-PdgSurf}. It is clear that the above expression for conjugate momenta to $q^{AB}$ and $l^{\mu}$ are identical to the corresponding expressions obtained in \cite{Parattu2015}. Thus the conjugate momentum $P_{\mu \nu}$, derived for general boundaries, is consistent with the corresponding ones for both the spacelike/timelike boundary as well as for the null boundary. 

A remark regarding the case of a non-vanishing reference connection is in order. One might recognize that the terms in \ref{WAC-MomRefConn} have the same structure as those in \ref{WAC-ConjMom1}. Now consider a situation where $\bar\Gamma{^\sigma}_{\mu \nu}$ is compatible with some metric $\bar{g}_{\mu \nu}$, such that the metrics ${g}_{\mu \nu}$ and $\bar{g}_{\mu \nu}$ (but not their derivatives) coincide at the boundary surface. Then the contribution of the reference connection to the conjugate momentum $P_{\mu \nu}$ may be expressed in terms of the extrinsic and mean curvatures associated with $\bar{g}_{\mu \nu}$ and will bear an identical geometric interpretation as the ones presented above.


%
%

%
%
\section{The gravitational energy-momentum tensor}\label{grav_em_tensor}

Here, we discuss the properties of the canonical energy-momentum tensor ${_{\rm E}}\Theta{_\nu}{^\mu}$, derived from the Weiss variation of the Einstein-Hilbert action with a boundary term, in some detail. In general, ${_{\rm E}}\Theta{^\mu}{_\beta}$ is non-vanishing, tensorial, and as we shall see shortly, satisfies a conservation law for flat reference geometries. Given the fact that this tensor is referred to as the energy-momentum tensor for the gravitational field, one might wonder whether it can serve as a definition for gravitational energy. However, ${_{\rm E}}\Theta{^\mu}{_\beta}$ is not unique, as it depends on the choice of the background geometry, and one can always choose a reference geometry to make ${_{\rm E}}\Theta{^\mu}{_\beta}$ vanish. It is evident that by employing the auxiliary connection formalism, we have traded off the background independence for a tensorial expression for energy and momentum of the gravitational field. Moreover, we shall see that even if one chooses a flat background geometry, different coordinate choices for the same can lead to different values for the gravitational energy-momentum tensor ${_{\rm E}}\Theta{^\mu}{_\beta}$; the non-covariance of the Einstein energy-momentum pseudotensor are now shifted to the ambiguities in the choice of the reference connection/background geometry. Of course, one might expect this, as equivalence principle arguments suggest that one cannot have a local definition for gravitational energy, as a point-like particle in free fall does not experience any gravitational field. In spite of this, ${_{\rm E}}\Theta{^\mu}{_\beta}$ is still of use in some specific contexts, or, can also be used as a dynamical definition for energy and hence we will pursue it further. 

We would also like to point out that the above expression for ${_{\rm E}}\Theta{^\mu}{_\beta}$ coincides with that of the energy-momentum pseudo-tensor for Einstein gravity, for vanishing reference connection. In which case, the quadrupole moment formula for gravitational radiation can be derived, following \cite{Pauli}.

\subsection{Conservation law}

The Einstein energy-momentum pseudotensor satisfies a nontensorial conservation law, so one might expect its tensorial ${_{\rm E}}\Theta{_\nu}{^\mu}$ to satisfy a tensorial conservation law. We show here that this is indeed the case if the reference geometry is flat, consistent with the result of \cite{Katzetal1997}. We consider a reference geometry with metric $\eta_{\mu \nu}$ and connection $\bar{\nabla}_\alpha \eta_{\mu \nu} = 0$. If the reference geometry is flat, then there exist coordinates such that $\bar{\Gamma}^\nu{_{\alpha \beta}}=0$, and it follows that the following quantity must vanish:
\begin{equation} \label{WAC-XiLDx}
  \begin{aligned}
  \tilde{\Xi}^\mu
  =&~
  \delta{^{\mu \alpha}}{_{\nu \sigma}} g^{\sigma \beta} \pounds_\xi \bar{\Gamma}^\nu{_{\alpha \beta}}.
  \end{aligned}
\end{equation}

\noindent As remarked earlier, there exists a class of covariantly constant vectors for a flat geometry; if we choose $\xi^\mu$ satisfying $\bar{\nabla}_{\nu} \xi^\mu=0$, then upon taking the divergence of \ref{WAC-KOri}, we obtain:
\begin{equation} \label{WAC-XiLD3Div}
  \begin{aligned}
  \nabla_\mu \left(\xi^\beta \left[2 G^{\mu}{_\beta} - {_{\rm E}}\Theta{^\mu}{_\beta}\right] \right) = 0 ,
  \end{aligned}
\end{equation}

\noindent where we have used $\sigma{^\mu}{_\beta}=0$, since $\bar{\nabla}_\mu$ is now torsion-free. Now one can show that for a covariantly constant vector (see derivation of \ref{GREC-DivXiT3B} in appendix):
\begin{equation}\label{WAC-DivXiT3B}
  \begin{aligned}
  \nabla_\sigma
  \left[
    \xi^\tau \, {S}{^\sigma}{_\tau}
  \right]
  = \, &
    \sqrt{\frac{\eta}{g}} \,
    \xi^\tau \,
    \bar{\nabla}_\sigma
    \left[
      \sqrt{\frac{g}{\eta}}
      {S}{^\sigma}{_\tau}
    \right]
  .
  \end{aligned}
\end{equation}

\noindent Combined with \ref{WAC-XiLD3Div}, and requiring that \ref{WAC-DivXiT3B} holds for all covariantly constant $\xi^\mu$,
\begin{equation}\label{WAC-dSconslaw}
  \begin{aligned}
    \bar{\nabla}_\sigma
    \left[
      \frac{\sqrt{-g}}{\sqrt{-\eta}}
      \left(2 G{^\sigma}{_\tau} - {_{\rm E}}\Theta{^\sigma}{_\tau}\right)
    \right] = 0
  .
  \end{aligned}
\end{equation}

\noindent We then find that the following quantity is divergence-free with respect to the connection of a flat metric $\eta_{\mu \nu}$:
\begin{equation}\label{GREC-dS}
  \begin{aligned}
  \Pi{^\sigma}{_\tau}
  =&~
  \frac{\sqrt{-g}}{\sqrt{-\eta}}
  \left(
  2 G{^\mu}{_\nu} - {_{\rm E}}\Theta{^\mu}{_\nu}
  \right)
  .
  \end{aligned}
\end{equation}

\noindent The form of $\Pi{^\sigma}{_\tau}$ is rather suggestive; upon replacing $G{^\mu}{_\nu}$ with the energy-momentum tensor for matter $G{^\mu}{_\nu}$, one may interpret ${_{\rm E}}\Theta{^\mu}{_\nu}$ as playing the role of a gravitational energy---in fact, this tensor has been of recent interest for characterizing the energy in gravitational radiation \cite{Cai2021}. The tensor $\Pi{^\sigma}{_\tau}$ is dependent on the coordinates for the flat reference, but even in the absence of a natural gauge choice, the quantity $\Pi{^\sigma}{_\tau}$ may be of utility as a dynamically conserved quantity. An interesting question, which we leave for future investigation, concerns the superpotential for $\Pi{^\sigma}{_\tau}$, which was obtained in \cite{Katzetal1997} (see Eq. (2.51) in that reference) for a Ricci-flat background. In particular, one might imagine that this potential is a tensorial generalization of the Freud superpotential \cite{Freud1939,Chen2018a,*Chen2018b}, and also ask whether such a superpotential yields a quasilocal energy equivalent to the Wang-Yau mass, as one might require \cite{Wang2008,Chen2018a,*Chen2018b}. Particularly interesting is the relationship between choice of reference connection in the approach presented here and the details of the embedding of the codimension two boundary surface into Minkowski space needed to construct the quasilocal energy.

\subsection{Harmonic coordinates}

A natural choice for the reference connection is then the flat Levi-Civita connection of Minkowski space, since the conservation law for $\Pi{^\sigma}{_\tau}$ applies with respect to the flat connection. However, this choice is still ambiguous without an appropriate coordinate condition on the metric $g_{\mu \nu}$. For instance, ${_{\rm E}}\Theta{^\mu}{_\beta}$ vanishes for Schwarzschild spacetime in the standard Schwarzschild coordinates but is nontrivial in Painlev{\'e}-Gullstrand coordinates---in the latter it is unclear whether the result is physically meaningful. In \cite{Nakanishi1986}, it was argued that under the Harmonic or De Donder coordinate condition (which corresponds to the condition $g^{\mu \nu} W{^\alpha}_{\mu \nu} = 0$ \cite{Bicak2005}), one obtains a physically meaningful result for the Schwarzschild metric. Though we find that there is indeed one Harmonic coordinate choice that yields a sensible result for the exterior of a Schwarzschild black hole, this is not the case in general.

In the simplest harmonic coordinate choice, the line element for the Schwarzschild metric of mass $m$ takes the form:
\begin{equation} \label{WAC-SchHarm}
  \begin{aligned}
 ds^2
  =& -\Phi_1 \, dt^2 + \Phi_2 \left(dx^2 + dy^2 + dz^2\right) \\
  & + \Phi_3 \left(x \, dx + y \, dy + z \, dz\right)^2,
  \end{aligned}
\end{equation}

\noindent where:
\begin{equation} \label{WAC-SchHarmPhis}
  \begin{aligned}
\Phi_1 & = \frac{\tilde{r}-m}{\tilde{r}+m}
    \qquad \Phi_2 = \frac{(\tilde{r}+m)^2}{\tilde{r}^2}
    \qquad \Phi_3 = \frac{m^2(\tilde{r}+m)}{\tilde{r}^4(\tilde{r}-m)}
  \end{aligned}
\end{equation}

\noindent with $\tilde{r}^2 = x^2 + y^2 + z^2$. The radial coordinate $\tilde{r}$ is related to the Schwarzschild radial coordinate by $r=\tilde{r}+m$; the horizon is located at $\tilde{r}=m$. In these coordinates, the energy density $e := {_{\rm E}}\Theta{^\mu}{_\nu} n_\mu \chi^\nu$ takes the form:
\begin{equation} \label{WAC-SchHarmEnergyDensity}
  \begin{aligned}
    e = -\frac{2 m^2 \sqrt{\tilde{r}^2-m^2}}{\tilde{r}^2 (m+\tilde{r})^3},
  \end{aligned}
\end{equation}

\noindent where $\chi=\partial/\partial t$ is a timelike Killing vector and $n=\chi/\sqrt{\Phi_1}$ is its normalized counterpart. Upon transforming to spherical coordinates, the volume element on a constant $t$ surface takes the form:
\begin{equation} \label{WAC-SchHarmVolelem}
  \begin{aligned}
    d\Sigma = \frac{(m+\tilde{r})^{5/2} \sin \theta }{\sqrt{\tilde{r}-m}},
  \end{aligned}
\end{equation}

\noindent and the indefinite integral of the energy density takes the form:
\begin{equation} \label{WAC-SchHarmEnergyIndefinite}
  \begin{aligned}
    E_I = \frac{m^2}{2 \tilde{r}} = \frac{m^2}{2 (r-m)}.
  \end{aligned}
\end{equation}

\noindent Integrating from the horizon, one has:
\begin{equation} \label{WAC-SchHarmEnergy}
  \begin{aligned}
    E = - \frac{m}{2} + \frac{m^2}{2 \tilde{r}} =  - \frac{m}{2} + \frac{m^2}{2 (r-m)},
  \end{aligned}
\end{equation}

\noindent and integrating out to infinity, one finds that $E \rightarrow -m/2$, which one might interpret as a gravitational binding energy.

Of course, the simple harmonic coordinate choice considered above is not unique. For the Schwarzschild spacetime, one can construct a more general relation between the harmonic coordinate radius $\tilde{r}$ and the areal radius $r$ \cite{Liu1998,Bicak2005}:
\begin{equation} \label{WAC-SchHarmRadiusRelation}
  \tilde{r}(r) =  c_1 (r-m)
                  +
                  c_2 \left\{(r-m) \, \ln \left[1-\frac{2 m}{r}\right] + 2 m \right\}
\end{equation}

\noindent Since $W{^\sigma}_{\mu \nu}$ is tensorial, one can compute its value in the Schwarzschild coordinates by using the above expression for $\tilde{r}(r)$ to transform the flat line element:
\begin{equation} \label{WAC-FlatLineElement}
  ds^2 = - dt^2 + d\tilde{r}^2 + \tilde{r}^2 d\Omega^2
\end{equation}

\noindent to the areal radius coordinates. The reference connection $\bar{\Gamma}{^\sigma}_{\mu \nu}$ may then be constructed from the corresponding metric. The indefinite energy integral takes the form:
\begin{equation} \label{WAC-SchHarmEnergyGeneral}
  \begin{aligned}
    E_I = \frac{m^2}{2(r - m)} - c_2 m^3 (\epsilon_1 + \epsilon_2) ,
  \end{aligned}
\end{equation}

\noindent where:
\begin{equation} \label{WAC-SchHarmEnergyGeneralII}
  \begin{aligned}
    \frac{1}{\epsilon_1} = \, & 2 c_2 m (m-r)
    + r (2 m-r) \left\{ c_1 + c_2 \ln \left[1-\frac{2 m}{r}\right] \right\}\\
    \frac{1}{\epsilon_2} = \, & (m-r)^2 \left(c_1 - \frac{2 c_2 m}{m-r}+ c_2 \ln \left(1-\frac{2 m}{r}\right)\right).
  \end{aligned}
\end{equation}

\noindent For $c_1 > 0$ and $c_2 > 0$, one finds that there exists a value $r>2$ such that $1/\epsilon_2 = 0$, leading to a divergent energy. One also finds that the energy $E_I$ changes sign on each side of the divergent point; the sign is indefinite. However, this divergent point can be pushed arbitrarily close to the horizon as one decreases the value for $c_2 > 0$. While the general Harmonic coordinate choice $c_2 > 0$ yields a pathological result for the energy obtained from ${_E}\Theta{^\mu}{_\beta}$ in black hole spacetimes, it is required for describing the exterior solution of spherically symmetric compact objects \cite{Liu1998,Bicak2005}. Moreover, one might interpret the pathologies in the energy as resulting from the fact that the coordinate relation in \ref{WAC-SchHarmRadiusRelation} is non-monotonic. Regardless, this demonstrates that while certain harmonic coordinate choices yield sensible expressions for the energy, other harmonic coordinate choices can yield pathological results. An important question for future investigation is whether one can always construct harmonic coordinates which yield well-behaved energy densities in generic vacuum spacetimes on and within a neighborhood of an initial time slice.


%
%

%
%
\section{Weiss variation in the matter sector}\label{weiss_matter}

For completeness, we work out the Weiss variation for a matter action of the form:
\begin{equation} \label{WAC-MatterAction}
S_{M} = \int_{M} d^4 x \, \sqrt{-g} \, \mathcal{L}(\varphi^\cdot,{\nabla}_\cdot\varphi^\cdot).
\end{equation}

\noindent where $\mathcal{L}$ is the matter Lagrangian density, which is assumed to depend only on the fields $\varphi^I$ and their derivatives ${\nabla}_\mu \varphi^I$. The variation of the action takes the form  (note $T^0_{\mu \nu}$ is \textit{not} the Hilbert energy-momentum tensor):
\begin{equation} \label{WAC-MatterActionVar}
\begin{aligned}
\Delta S_{M} &= \int_{M} d^4 x \, \sqrt{-g} \biggl[ \mathcal{E}_I \delta \varphi^I - \tfrac{1}{2}T^0_{\mu \nu} \delta g^{\mu \nu} + S^{\mu \nu}{_\sigma} \delta {\Gamma}{^\sigma}_{\mu \nu} \biggr]\\
& \quad + \int_{\partial M} d\Sigma_\mu \biggl[ \mathcal{L} \delta x^\mu + P{_I}{^\mu} \delta \varphi^I \biggr],
\end{aligned}
\end{equation}

\noindent where:
\begin{equation} \label{WAC-ELoperator}
\mathcal{E}_I := \frac{\partial \mathcal{L}}{\partial \varphi^I} -  {\nabla}_\mu P{_I}{^\mu}
\end{equation}
\begin{equation} \label{WAC-ELoperatorII}
P{_I}{^\mu} := \frac{\partial \mathcal{L}}{{\nabla}_\mu \partial \varphi^I}.
\end{equation}

\noindent The variation of the Levi-Civita connection coefficients can be expressed in terms of the covariant derivatives of the inverse metric variations:
\begin{equation} \label{ChristoffelSymbolVariationInv}
\delta {\Gamma}^{\sigma \mu \nu}=-\frac{1}{2}({\nabla}^\mu \delta g^{\tau \nu}+{\nabla}^\nu \delta g^{\mu \tau}-{\nabla}^\tau \delta g^{\mu \nu}).
\end{equation}

\noindent One may then write:
\begin{equation} \label{ChristoffelSymbolVariationS}
\begin{aligned}
S^{\mu \nu}{_\sigma} \delta {\Gamma}{^\sigma}_{\mu \nu}
&=
s_{\mu \nu \sigma} \, {\nabla}^\sigma \delta g^{\mu \nu},
\end{aligned}
\end{equation}

\noindent where:
\begin{equation} \label{SpinVar}
\begin{aligned}
s_{\mu \nu \sigma}
&:=
\frac{1}{2}
\biggl[
S_{\mu \nu \sigma}-S_{\sigma \nu \mu}-S_{\mu \sigma \nu}
\biggr].
\end{aligned}
\end{equation}

\noindent Since $S_{\mu \nu \sigma}$ is symmetric in the first two indices, it follows that $s_{\mu \nu \sigma}$ must be as well. It should be mentioned that for scalar fields and gauge fields, the tensor $S_{\mu \nu \sigma}$ is identically zero, and for Dirac spinor fields, $S_{\mu \nu \sigma}$ (and $s_{\mu \nu \sigma}$) is completely antisymmetric.

After an integration by parts, the variation of the matter action then takes the form:
\begin{equation} \label{WAC-MatterActionVarBOld}
\begin{aligned}
\Delta S_{M} &= \int_{M} d^4 x \, \sqrt{-g} \biggl[ \mathcal{E}_I \delta \varphi^I - \tfrac{1}{2}T_{\mu \nu} \delta g^{\mu \nu}
\biggr]\\
& \quad + \int_{\partial M} d^3 \Sigma_\sigma \biggl[ \mathcal{L} \delta x^\sigma + P{_I}{^\sigma} \delta \varphi^I
+ g^{\sigma \tau} \,
s_{\mu \nu \tau}\delta g^{\mu \nu}\biggr].
\end{aligned}
\end{equation}

\noindent It may be helpful at this point to examine the form of the Lie derivative in greater detail. For tensorial fields, the Lie derivative takes the form:
\begin{equation} \label{LieDerivTensor}
\begin{aligned}
\pounds_V T{^{\mu_1 ... \mu_m }}{_{\nu_1 ... \nu_n}}
:=&
V^\sigma {\nabla}_\sigma T{^{\mu_1 ... \mu_m }}{_{\nu_1 ... \nu_n}} \\
& - T{^{\sigma ... \mu_m }}{_{\nu_1 ... \nu_n}} {\nabla}_\sigma V{^{\mu_1}} - ... \\
& - T{^{\mu_1 ... \sigma }}{_{\nu_1 ... \nu_n}} {\nabla}_\sigma V{^{\mu_m}}\\
& + T{^{\mu_1 ... \mu_m }}{_{\sigma ... \nu_n}} {\nabla}{_{\nu_1}} V^\sigma + ... \\
& + T{^{\mu_1 ... \mu_m }}{_{\nu_1 ... \sigma}} {\nabla}{_{\nu_n}} V^\sigma,
\end{aligned}
\end{equation}

\noindent and for spinor fields (here working in the orthonormal basis):
\begin{equation} \label{LieDerivSpinor}
\begin{aligned}
\pounds_V \psi
:=&
V^a {\nabla}_a \psi - \tfrac{1}{2} {\nabla}_a V_b \, \sigma^{ab}  \psi \\
\pounds_V \tilde{\psi}
:=&
V^a {\nabla}_a \tilde{\psi} + \tfrac{1}{2} {\nabla}_a V_b \, \tilde{\psi} \sigma^{ab} ,
\end{aligned}
\end{equation}

\noindent where $\sigma^{ab}$ are generators of the Lorentz transformation. In each case, we note that there is a part proportional to $V$, and a part that depends on the derivatives of $V$, so one may write the Lie derivative for the fields $\varphi^I$ compactly as:
\begin{equation} \label{LieDerivGeneral}
\begin{aligned}
\pounds_V \varphi^I
:=&
V^\sigma {\nabla}_\sigma \varphi^I + w{_J}^{I} \varphi^J,
\end{aligned}
\end{equation}

\noindent where $w{_J}^{I}$ depends on the derivative of $V$. The boundary variation of the field $\varphi^I$ takes the form:
\begin{equation} \label{bdvfield}
\begin{aligned}
\Delta \varphi^I
:=& \delta \varphi^I + \pounds_{\Delta \lambda \xi} \varphi^I \\
=& \delta \varphi^I + \Delta \lambda \xi^\sigma {\nabla}_\sigma \varphi^I + w{_J}^{I} \varphi^J .
\end{aligned}
\end{equation}

The variation of the action becomes (with ${_M}\Theta{^\sigma}{_\mu}$ being the canonical energy-momentum tensor for the matter fields):
\begin{equation} \label{WAC-MatterActionVarB}
\begin{aligned}
\Delta S_{M} &= \int_{M} d^4 x \, \sqrt{-g} \biggl[ \mathcal{E}_I \delta \varphi^I - \tfrac{1}{2}T_{\mu \nu} \delta g^{\mu \nu}
\biggr]\\
& \quad + \int_{\partial M} d^3 \Sigma_\sigma \biggl[P{_I}{^\sigma} \Delta \varphi^I + g^{\sigma \tau} \, s_{\mu \nu \tau} \Delta g^{\mu \nu}\\
& \qquad \qquad \qquad \quad
\Delta \lambda \xi^\mu \biggl\{{_M}\Theta{^\sigma}{_\mu} + {\nabla}^{\nu} s_{\mu \nu}{^\sigma} - {\nabla}^\nu Y_{\mu \nu}{^\sigma} \biggr\}\\
& \qquad \qquad \qquad \quad
+ {\nabla}_{\nu} \left[ \Delta \lambda \xi^\mu  ( s_{\mu}{^{\nu \sigma}} - Y_{\mu}{^{\nu \sigma}})\right]
\biggr].
\end{aligned}
\end{equation}

\noindent where $Y_{\mu \nu}{^\sigma}$ is defined such that:
\begin{equation} \label{WAC-MatterActionVarBmom}
\begin{aligned}
P{_I}{^\sigma} w{_J}^{I} \varphi^J = {\nabla}^\nu (\Delta \lambda \, \xi^\mu \, Y_{\mu \nu}{^\sigma})- \Delta \lambda \, \xi^\mu \, {\nabla}^\nu Y_{\mu \nu}{^\sigma} ,
\end{aligned}
\end{equation}

\noindent
One might then imagine the following quantity to be the potential used to obtain the Belinfante-Rosenfeld tensor \cite{Belinfante1940,*Rosenfeld1940}:
\begin{equation} \label{WAC-BRTensor}
\tilde{s}_{\mu \nu \sigma} = s_{\mu \nu \sigma} - Y_{\mu \nu \sigma}.
\end{equation}

\noindent It is straightforward to show that the last term in \ref{WAC-MatterActionVarB} can be converted to a bulk integral that identically vanishes if $\tilde{s}_{\mu \nu \sigma}$ is antisymmetric in the last two indices $\nu$ and $\sigma$. In particular, one can show that if $\tilde{s}_{\mu \nu \sigma}=-\tilde{s}_{\mu \sigma \nu}$, then one may write (with $a^{[\sigma \nu]}=\Delta \lambda \xi^\mu \tilde{s}_{\mu}{^{\nu \sigma}}$):
\begin{equation} \label{WAC-MatterActionVarBLT}
\begin{aligned}
\int_{\partial M} d^3 \Sigma_\sigma \,{\nabla}_{\nu} a^{[\sigma \nu]}
 = \int_{M} d^4 x \sqrt{-g}  \left({\nabla}_\sigma {\nabla}_{\nu} a^{[\sigma \nu]} \right).
\end{aligned}
\end{equation}

\noindent Since $a^{[\sigma \nu]} = - a^{[\nu \sigma]}$, it is straightforward to show that ${\nabla}_\sigma {\nabla}_{\nu} a^{[\sigma \nu]}={R}_{\sigma \nu} a^{[\sigma \nu]}=0$.

It is perhaps appropriate to compute $Y_{\mu \nu}{^\sigma}$ for a few explicit examples. We first consider a gauge field $A_\mu$ (with gauge index suppressed). The field strength tensor takes the form:
\begin{equation} \label{WAC-FieldStrengthTensor}
F_{\mu \nu} = {\nabla}_{\mu} A_{\nu} - {\nabla}_{\nu} A_{\mu} + c f A_\mu A_\nu,
\end{equation}

\noindent where $c$ is a coupling constant and $f$ represents the structure constants (again, gauge indices are suppressed). The Lagrangian density for gauge fields has the form:
\begin{equation} \label{WAC-GaugeFieldLagrangian}
\mathcal{L}_{gf} = -\frac{1}{4} F^{\mu \nu} F_{\mu \nu}.
\end{equation}

\noindent The polymomentum takes the form:
\begin{equation} \label{WAC-GaugeFieldPolymomentum}
P^{\mu \nu} = - F^{\mu \nu}.
\end{equation}

\noindent The equivalent expression for \ref{WAC-MatterActionVarB} in this case becomes:
\begin{equation} \label{WAC-GaugeFieldPolymomentumY}
\begin{aligned}
P^{\mu \nu} \, {\nabla}_\nu \left(\Delta \lambda \, \xi^\sigma \right) A_\sigma
= \,
& {\nabla}_\nu \left(\xi^\sigma \, \Delta \lambda \, P^{\mu \nu} \, A_\sigma \right) \\
& - \Delta \lambda \, \xi^\sigma \, {\nabla}_\nu \left(P^{\mu \nu} \, A_\sigma \right)
\end{aligned}
\end{equation}

\noindent which upon comparison with \ref{WAC-MatterActionVarBmom}, allows for the identification of $Y_{\mu \nu}{^\sigma}$:
\begin{equation} \label{WAC-Ygauge}
\begin{aligned}
Y_{\mu \nu \sigma} = P_{\sigma \nu} \, A_\mu = - F_{\sigma \nu} \, A_\mu
\end{aligned}
\end{equation}

\noindent and it follows that $Y_{\mu \nu \sigma} = - Y_{\mu \sigma \nu}$. \ref{WAC-Ygauge} is in fact the form for $Y_{\mu \nu \sigma}$ needed to construct the Belinfante-Rosenfeld tensor.

Now we turn to spinor fields. The free massless spinor Lagrangian takes the form:
\begin{equation} \label{WAC-SpinorLagrangian}
\mathcal{L}_{sp} = \frac{i}{2}(\bar{\psi} \gamma^a {\nabla}_a \psi - {\nabla}_a \bar{\psi} \gamma^a  \psi ),
\end{equation}

\noindent where $c.c.$ denotes complex conjugation (to ensure that the resulting action is ultimately real). The polymomenta for the Lagrangian take the form (where $\gamma^\mu :=e_a^\mu \, \gamma^a$):
\begin{equation} \label{3F-DiracLagrangianDerivatives}
\begin{array}{cc}
\begin{aligned}
\frac{\partial \mathcal{L}_{sp}}{\partial {\nabla}_\mu \bar{\psi}}&=\bar{P}^\mu=-i \> \gamma^\mu \>  {\psi}, \\
\end{aligned}
\>\>\>\>\>\>\>\> & \>\>\>\>\>\>\>\>
\begin{aligned}
\frac{\partial \mathcal{L}_{sp}}{\partial {\nabla}_\mu\psi}&={P}^\mu=i \> \bar{\psi} \, \gamma^\mu .
\end{aligned}
\end{array}
\end{equation}

\noindent For spinors, the equivalent for \ref{WAC-MatterActionVarB} becomes:
\begin{equation} \label{WAC-SpinorPolymomentumY}
\begin{aligned}
P{^\sigma} {\nabla}_\mu \left(\Delta \lambda \, \xi_\nu \right) \sigma^{\mu \nu} \psi =
&
{\nabla}_\mu \left( \Delta \lambda \, \xi_\nu \, P{^\sigma} \, \sigma^{\mu \nu} \psi \right) \\
& - \Delta \lambda \, \xi_\nu \, {\nabla}_\mu \left(P{^\sigma} \, \sigma^{\mu \nu} \psi \right) \\
{\nabla}_\mu \left(\Delta \lambda \, \xi_\nu \right) \bar{\psi} \, \sigma^{\mu \nu} \, \bar{P}{^\sigma}  =
&
{\nabla}_\mu \left( \Delta \lambda \, \xi_\nu \, \bar{\psi} \, \sigma^{\mu \nu} \, \bar{P}{^\sigma} \right) \\
& - \Delta \lambda \, \xi_\nu \, {\nabla}_\mu \left( \bar{\psi} \, \sigma^{\mu \nu} \, \bar{P}{^\sigma} \right),
\end{aligned}
\end{equation}

\noindent from which one may upon comparison with \ref{WAC-MatterActionVarBmom} identify $Y^{\mu \nu \sigma}$:
\begin{equation} \label{WAC-Yspinor}
\begin{aligned}
Y^{\mu \nu \sigma} &= - \tfrac{1}{2}P^{\sigma} \, \sigma^{\nu \mu} \, \psi + \tfrac{1}{2} \bar{\psi} \, \sigma^{\nu \mu} \, \bar{P}^{\sigma} \\
&= - \frac{i}{2} \, \bar{\psi} \left( \gamma^\sigma \, \sigma^{\nu \mu} + \sigma^{\nu \mu} \, \gamma^\sigma \right) \psi .
\end{aligned}
\end{equation}

\noindent The above can be can shown to be completely antisymmetric. We have thus shown that for the types of fields appearing in the standard model, $Y^{\mu \nu \sigma}$ is antisymmetric in the last two indices and the tensor
\begin{equation} \label{WAC-BRTensorII}
\begin{aligned}
\bar{T}{^\sigma}{_\mu} = {_M}\Theta{^\sigma}{_\mu} + {\nabla}^{\nu} \left( s_{\mu \nu}{^\sigma} - Y_{\mu \nu}{^\sigma} \right)
\end{aligned}
\end{equation}

\noindent is equivalent to the Belinfante-Rosenfeld tensor if $s_{\mu \nu}{^\sigma} = 0$. Now $s_{\mu \nu}{^\sigma}$ has been kept for completeness, but for gauge fields, it does not appear since the connection coefficients (assuming they are torsion-free) cancel. For spinor fields, $s_{\mu \nu \sigma}$ is completely antisymmetric, and since it is contracted with the metric variations $\delta g^{\mu \nu} = \delta g^{\nu \mu}$, $s_{\mu \nu \sigma}$ does not contribute to the variation. However, for more general matter models which have a nontrivial coupling to the the connection, $s_{\mu \nu \sigma}$ may contribute nontrivially to the boundary variations.

Combining the results, we find that for $S=S_{\rm gW}+S_{\rm M}$, the Weiss variation of the action takes the form:
\begin{equation} \label{WAC-ACFActionVarFullMat}
\begin{aligned}
\Delta S =&\int_{U} d^4 x  \sqrt{-g} \, \left[\frac{1}{2 \kappa} \left(G_{\mu \nu} - \kappa T_{\mu \nu} \right) \delta g^{\mu \nu} + \mathcal{E}_I \delta \varphi^I \right] \\
& + \int_{\partial U} d\Sigma_\mu \biggl[\frac{1}{2 \kappa}\delta\bar{\Gamma}^\mu + P{_I}{^\mu} \Delta \varphi^I \\
& + \left\{\frac{1}{2 \kappa} P{^\mu}_{\alpha \beta} + g^{\mu \tau} s_{\alpha \beta \tau} \right\} \Delta g^{\alpha \beta} \\
& + \frac{\Delta \lambda}{2 \kappa} \biggl\{ [ 2 \kappa \bar{T}{^\mu}{_\beta} - {_E}\Theta{^\mu}{_\beta} + \sigma{^\mu}{_\beta} ]\xi^\beta
    + \Sigma^{\mu \alpha}{_\beta} \bar{\nabla}_\alpha \xi^\beta \biggr\} \biggr].
\end{aligned}
\end{equation}

\noindent By comparison with the matter field results, one might expect the last term involving $\Sigma^{\mu \alpha}{_\beta}$ to yield a Belinfante-Rosenfeld type correction to ${_E}\Theta{^\mu}{_\beta}$ \cite{Katzetal1997}. However, the main obstacle to such an interpretation is that $\Sigma^{\mu \alpha}{_\beta}$ is not antisymmetric in the indices $\mu \alpha$, so in general, one obtains an additional term that cannot be straightforwardly eliminated with the procedure described in the paragraph containing \ref{WAC-MatterActionVarBLT}. Belinfante-Rosenfeld corrections for ${_E}\Theta{^\mu}{_\beta}$ have been obtained elsewhere \cite{Papapetrou1948,Katzetal1997,PetrovKatz2002,PetrovPitts2020}; an interesting question for future investigation is whether one can nontrivially\footnote{For a covariantly constant vector $\xi^\mu$ with respect to a flat connection $\bar{\nabla}_\alpha$, one has $\bar{\nabla}_\alpha \xi^\beta = 0$. In this case, one can trivially subtract $\Sigma^{(\mu \alpha)}{_\beta}\bar{\nabla}_\alpha \xi^\beta =0$ from the boundary term in \ref{WAC-ACFActionVarFullMat} and perform an integration by parts (eliminating a term using \ref{WAC-MatterActionVarBLT} and the identity ${\bar\nabla}_\sigma {\bar\nabla}_{\nu} a^{[\sigma \nu]}=\bar{R}_{\sigma \nu} a^{[\sigma \nu]}=0$) to obtain the boundary integrand $({\Delta \lambda}/{2 \kappa})\left[ 2 \kappa \bar{T}{^\mu}{_\beta}- {_E}\Theta{^\mu}{_\beta}+\sqrt{\eta/g}\bar{\nabla}_\alpha(\sqrt{g/\eta} \, \Sigma^{[\mu \alpha]}{_\beta})\right] \xi^\beta$.} recover the same Belinfante-Rosenfeld energy-momentum tensor from the last term in \ref{WAC-ACFActionVarFullMat} under the appropriate conditions.


%
%

%
%
\section{Einstein-Schr{\"o}dinger equation}\label{einstein_schrodinger}

The Hamilton-Jacobi equation for general relativity has been studied in detail since the 1960s \cite{Peres1962,*Bergmann1966,*Komar1967,*Komar1968,*Gerlach1969,*Komar1970,*Komar1971,*Goldberg1994,*Goldberg1996,*Rovelli2004,*Salisbury2015,Wheeler1968} (see also \cite{Salisbury2021} for a recent historical review), and was recently derived via the Weiss variation in \cite{FengMatzner2018}. A natural question is whether the Weiss variation may also be used to derive a Schr{\"o}dinger equation for general relativity. In fact, the Weiss variation provides a relatively straightforward (albeit formal) derivation of the Wheeler-DeWitt equation \cite{DeWitt1967,Wheeler1968} from the path integral from quantum general relativity \cite{FengMatzner2017}. Here, we revisit the derivation for the Weiss variation in the case of vacuum gravity, this time with the variation as in \ref{WAC-ACFActionVarFullRW5}. The formal amplitude $\mathscr{A}$ for vacuum gravity takes the form:
\begin{equation} \label{WAC-FunctionalIntegral}
  \mathscr{A}[\mathring{g}^{\cdot \cdot}] = \int \mathscr{D} g \, e^{(i/\hbar) S_{gW}[g^{\cdot \cdot}]},
\end{equation}

\noindent where $\mathring{g}^{\cdot \cdot}:=g^{\cdot \cdot}|_{\partial M}$ denotes the boundary value of $g^{\alpha \beta}$, $\int \mathscr{D} g$ formally represents a functional integral over $g^{\alpha \beta}$, with the boundary values $\mathring{g}^{\cdot \cdot}$ held fixed under functional integration, and the dependence on the reference connection $\bar{\Gamma}{^\mu}_{\alpha \beta}$ has been suppressed. If functional integration is formally invariant under field redefinitions (for instance $g^{\alpha \beta}\rightarrow g^{\alpha \beta} + \delta g^{\alpha \beta}$), one may recover the functional Ehrenfest theorem:
\begin{equation} \label{WAC-FunctionalEhrenfest}
  \int \mathscr{D} g \, \left(\sqrt{-g} \, G_{\mu \nu} \right) e^{(i/\hbar) S_{gW}[g^{\cdot \cdot}]} = 0 .
\end{equation}

\noindent The variation of the amplitude takes the form:
\begin{equation} \label{WAC-FunctionalIntegralVar}
  \Delta \mathscr{A}[\mathring{g}] = \frac{i}{\hbar} \int \mathscr{D} g \, \Delta S_{gW} \, e^{(i/\hbar) S_{gW}[g^{\cdot \cdot}]}.
\end{equation}

\noindent From the functional Ehrenfest theorem, the bulk term in $\Delta S_{gW}$ vanishes within the functional integral, and the variation of the amplitude will consist of boundary terms. Now assume the existence of a flat reference metric associated Levi-Civita connection $\bar{\Gamma}{^\mu}_{\alpha \beta}$. Under boundary displacements induced by a vector $\xi^\mu$ that is covariantly constant with respect to the reference geometry, the variation of the amplitude may be written as:
\begin{equation} \label{WAC-FunctionalIntegralVarBdy}
  \begin{aligned}
  \Delta \mathscr{A}[\mathring{g}] =&\frac{i}{2 \kappa \hbar} \biggl\{  \int_{\partial M} d\underline{\Sigma} \biggl[ \hat P_{\alpha \beta} \Delta g^{\alpha \beta}
  -
  \hat{\mathcal{H}}{_\nu} \, \xi^\nu \, \Delta \lambda
  \biggr]
  \biggr\},
  \end{aligned}
\end{equation}

\noindent where the following have been defined ($n_\mu$ being the dual to the normal vector of the boundary):
\begin{equation} \label{WAC-FunctionalIntegralVarBdyDefs}
  \begin{aligned}
  \hat P_{\alpha \beta} &:= \int \mathscr{D} g \, \sqrt{-\mathring{g}} \, n_\mu \, P{^\mu}_{\alpha \beta} \, e^{(i/\hbar) S_{gW}[g^{\cdot \cdot}]} \\
  \hat{\mathcal{H}}{_\nu} &:= \int \mathscr{D} g \, \sqrt{-\mathring{g}} \, n_\mu \, {_E}\Theta{^\mu}{_\nu} \, e^{(i/\hbar) S_{gW}[g^{\cdot \cdot}]}.
  \end{aligned}
\end{equation}

\noindent Note that since the functional integral is over $g^{\alpha \beta}$ fixed to $\mathring{g}$ at the boundary, one may pull out the boundary integrals from within the functional integral, as the boundary volume element is effectively constant with respect to the functional integral. With regard to the action on $\mathscr{A}[\mathring{g}^{\cdot \cdot}]$, $\hat P_{\alpha \beta}$ corresponds to the functional derivative of $\mathscr{A}[\mathring{g}^{\cdot \cdot}]$ with respect to $g^{\alpha \beta}$, and the operator $\hat g^{\alpha \beta}$ corresponds to multiplication by the boundary value:
\begin{equation} \label{WAC-FieldOperator}
  \hat g^{\alpha \beta} \mathscr{A}[\mathring{g}^{\cdot \cdot}] = \mathring{g}^{\alpha \beta} \mathscr{A}[\mathring{g}^{\cdot \cdot}],
\end{equation}

\noindent since the boundary values $\mathring{g}^{\alpha \beta}$ are held fixed in the functional integral and can be passed through the functional integration.

Now consider the case where the displaced portion of $\partial M$ is a spacelike surface (so that $n^\mu$ is timelike) and $\xi^\mu$ is timelike. One might, for instance consider $M=R \times S^3$ with $\partial M$ being the union of Cauchy surfaces, with one Cauchy surface displaced. One can in principle define a Hamiltonian operator,
\begin{equation} \label{WAC-HamiltonianOperator}
  \begin{aligned}
  \hat H  &:= \frac{1}{2 \kappa} \int_{\partial M} d\underline{\Sigma} \, \hat{\mathcal{H}}{_\nu} \, \xi^\nu ,
  \end{aligned}
\end{equation}

\noindent which by \ref{WAC-FunctionalIntegralVarBdy} is related to the derivative of the amplitude according to:
\begin{equation} \label{WAC-EinsteinSchrodinger}
  \begin{aligned}
  -i \hbar \frac{\partial \mathscr{A}}{\partial \lambda} = \hat H .
  \end{aligned}
\end{equation}

\noindent If $\hat{\mathcal{H}}{_\nu}$ can be written\footnote{This is a nontrivial matter, as the operator $\hat P_{\alpha \beta}$ corresponds to a functional derivative operator acting on the amplitude. Multiple factors of $P_{\alpha \beta}$ in the classical Hamiltonian correspond to multiple functional derivatives of the amplitude, which requires regularization \cite{DeWitt1967,Kiefer2012}. The regularization of such operators is an open issue \cite{Kiefer2020}; see \cite{FengWdW2018} for a recent proposal.} in terms of the operators $\hat P_{\alpha \beta}$ and $\hat g^{\alpha \beta}$, then \ref{WAC-EinsteinSchrodinger} becomes a Schr{\"o}dinger equation. Since the classical Hamiltonian is generally nonvanishing, this result at first sight appears to offer a resolution to the problem of time. However, the cost is a loss of uniqueness in the definition of the Hamiltonian operator $\hat H$, as it depends on the choice of coordinates on a flat reference geometry. One must also confront the potential pathologies that may arise with a poor choice of coordinates; as shown earlier, the classical Hamiltonian can change sign and diverge for certain coordinate coordinate choices. Again, these difficulties might be expected in light of the general expectation that a physically meaningful local definition for gravitational energy does not exist, in which case the problem of time persists. Moreover, \ref{WAC-EinsteinSchrodinger} obscures the role of the local Hamiltonian constraint (the Wheeler-DeWitt Equation), which should hold as a consequence of the Ehrenfest theorem. One might for these reasons expect the Hamiltonian operator $\hat H$ to ultimately reduce to a quasilocal quantity (in the sense that the right-hand side of \ref{WAC-HamiltonianOperator} is a surface integral), in line with expectations from the earlier discussion regarding the superpotential for $\Pi{^\sigma}{_\mu}$ and previous results \cite{FengMatzner2017,Rosabal2021,*Hayward1992}. Nonetheless, this result may still be a useful tool for understanding the relationship between reparameterization invariance and the problem of time, as well as a practical tool for characterizing time evolution in quantum general relativity (for instance in perturbation theory and for gravitational systems confined to finite regions of space).


%
%

%
%
\section{Discussion and summary}
We have performed the Weiss variation for general relativity using a covariant boundary term found in \cite{IsenbergNester1980,Nester1982,Lynden-Belletal1995,Harada2020CIGA,BeltranJimenez2019}, constructed with an auxiliary reference connection $\bar{\Gamma}{^\sigma}_{\mu \nu}$, which does not appear in the bulk action. The resulting Weiss variation is valid for general boundary surfaces and reduces to the expected expressions for spacelike/timelike as well as for null boundaries in appropriate gauges, extending the earlier result in \cite{FengMatzner2018,Parattu2015}, which was only valid for spacelike/timelike or, for null surfaces. In particular, we have worked out the conjugate momenta for both null and non-null surfaces, which agrees with the earlier derived results. Furthermore, we have argued that upon discarding certain surface terms, there is a part of the Weiss variation, explicitly dependent on boundary displacements (in particular, the parameter $\Delta \lambda$), that consists of a combination of constraints on the boundary, as expected from \cite{FengMatzner2018}.

It turns out that the boundary displacement terms can be rewritten, yielding a canonical gravitational energy-momentum tensor that generalizes the Einstein pseudo-tensor. Such a tensor was first discussed in \cite{Rosen1940} and \cite{Papapetrou1948}, and studied further in \cite{Lynden-Belletal1995,Katzetal1997}; we note that the same tensor has been of recent interest for characterizing the energy content of gravitational radiation \cite{Cai2021}. We discuss the properties of this gravitational energy-momentum tensor, in particular demonstrating that it satisfies a conservation law when a flat reference geometry is chosen (restricting to domains over which such a geometry exists), in line with \cite{Katzetal1997}. Upon examining the resulting energy expressions in the Schwarzschild geometry we find that, while one obtains a reasonable expression for a particular Harmonic coordinate choice, the energy expressions are in general highly dependent on the choice of coordinates on the reference geometry and can exhibit pathological behavior for certain coordinate choices. This indicates that despite the fact that while one has a tensorial expression for the canonical energy-momentum for gravity, the ambiguities of the Einstein pseudo-tensor is present in its tensorial generalization (the latter inheriting ambiguities from the choice of auxiliary connection). Moreover, the description is no longer background independent --- in effect, one trades background independence for a tensorial energy-momentum tensor. However, the tensorial expression may be preferable to the pseudo-tensorial expression, and it may still be possible to obtain a unique quasi-local energy from the superpotential for the canonical energy-momentum tensor for gravity, as in the pseudo-tensor case \cite{Chen2018a,*Chen2018b}; whether the superpotential in \cite{Katzetal1997} yields the same quasilocal energy is an interesting question which we leave for future work.

We extended the Weiss variation to the case where matter fields are also included. In this case, we find that Belinfante-Rosenfeld corrections for the matter fields appears in the boundary variations. Upon comparison with the gravitational case, a term playing a similar role to those that give rise to the Belinfante-Rosenfeld corrections (that involving the tensor $\Sigma^{\mu\alpha}_\beta$) also appears, but the term lacks a required anti-symmetry. Additional conditions are needed to recover the same Belinfante-Rosenfeld energy momentum tensor obtained elsewhere \cite{Papapetrou1948,Katzetal1997,PetrovKatz2002,PetrovPitts2020}; we leave for future investigation the problem of non-trivially obtaining the symmetrization from our boundary term.

In quantum general relativity, the Weiss variation facilitates the formal derivation of a gravitational Schr{\"o}dinger equation from path integral expressions for the amplitude. The Hamiltonian operator does not in general vanish, suggesting nontrivial time evolution for quantum states. However, the Hamiltonian operator suffers from the aforementioned ambiguities for the classical Hamiltonian, and one might expect the amplitudes to also be subject to a local Hamiltonian constraint (Wheeler-DeWitt Equation). These considerations lead one to expect that the Hamiltonian operator is ultimately quasi-local in nature, in line with expectations arising from the classical analysis and earlier work \cite{FengMatzner2017,Rosabal2021,*Hayward1992}; whether this is the case is an interesting matter for future study. That said, the gravitational Schr{\"o}dinger equation may still serve as a useful tool for studying time evolution in quantum general relativity for systems confined to finite regions of space.

The utility of the auxiliary connection formalism we have employed in obtaining our expressions for the Weiss variation arises from the fact that one can maintain covariance without restricting to boundaries that are strictly timelike, spacelike (as is the case with the Gibbons-Hawking-York boundary term) or, null for \cite{Parattu2015}. This work is more in the spirit of \cite{Parattu:2016trq}. The cost of using this formalism is the ambiguity in the choice of the reference connection $\bar{\Gamma}{^\mu}_{\alpha \beta}$, which leads to ambiguities in the canonical energy-momentum tensor. Despite this, the expressions that we have obtained for the Weiss variation may be of practical use in certain contexts, such as in the perturbation theory and the problem of determining the energy content of gravitational radiation. We also expect the results and methods described here to be useful for obtaining and studying the Hamiltonian formulation of general relativity in different variables and in complicated modified gravity theories, along the lines of \cite{Chakraborty:2017zep,Chakraborty:2018dvi}.

Having elaborated on our findings above, below we provide an executive summary of our main results: 
\begin{itemize}
\item We extend the Weiss variation for the Einstein-Hilbert action (with a boundary term) in \cite{FengMatzner2018} for spacelike/timelike boundary surfaces to general boundary surfaces. 
\item We extend the energy-momentum tensor of \cite{Lynden-Belletal1995} to general reference connections (as opposed to only Levi-Civita connections), and identify the trade offs one needs to perform, in order to construct a tensorial counterpart for the Einstein energy-momentum pseudo-tensor. 
\item We obtain the Belinfante-Rosenfeld corrections to the energy-momentum tensor for matter fields via Weiss variations, and discuss the difficulties in doing the same for the gravitational energy-momentum tensor. 
\item We revisit and extend the work of \cite{FengMatzner2017}, by deriving a formal Schr{\"o}dinger equation from the path integral, albeit with the caveats discussed at length in \ref{einstein_schrodinger} and the preceding paragraph.
\end{itemize}

This work will find applications in different contexts, e.g., --- (i) whether the ``Complexity=Action" conjecture holds true even when the boundary term involving reference connection is added needs to be demonstrated, (ii) whether there exist any possible thermodynamical interpretation for the gravitational energy-momentum tensor involving reference connection, (iii) if it is possible to derive a first law of thermodynamics using the full Noether current $J^{\mu}$, involving reference connection. We leave these for the future. 


%
%


\begin{acknowledgments}
Research of S.C. is funded by the INSPIRE Faculty fellowship from the DST, Government of India (Reg. No. DST/INSPIRE/04/2018/000893) and by the Start-Up Research Grant from SERB, DST, Government of India (Reg. No. SRG/2020/000409). J.C.F. is grateful for helpful discussions with Vincenzo Vitagliano, Stefano Vignolo, Sante Carloni, and Richard Matzner. J.C.F. also thanks DIME at the University of Genoa for hosting a visit during which part of this work was performed, and acknowledges financial support from FCT - Funda\c c\~ao para a Ci\^encia e Tecnologia of Portugal Grant No.~PTDC/MAT-APL/30043/2017 and Project No.~UIDB/00099/2020. Some of the calculations were performed using the xAct package \cite{xActbib} for \textit{Mathematica}. This manuscript has no available data.
\end{acknowledgments}


%
%

%
%
\appendix
\labelformat{section}{Appendix #1} 
\labelformat{subsection}{Appendix #1}

\section{Variation of bulk integrals}
Here, we work out the variation of bulk integrals when boundary displacements are included. Consider a bulk integral of the form:
\begin{equation} \label{WAC-BulkInt}
I = \int_U d^4 x \, F .
\end{equation}

\noindent The variation of the integral takes the form:
\begin{equation} \label{WAC-BulkIntVar0}
\begin{aligned}
\delta I
&= \int_{U^\prime} d^4 x \, F^\prime - \int_U d^4 x \, F - \mathcal{O}(2) \\
&= \int_U d^4 x \, \delta F + \int_{\partial U} d \underline{\Sigma}_\mu \, \delta x^\mu \, F .
\end{aligned}
\end{equation}

\noindent The second integral in the last equality describes the contribution from the boundary variation; one can see this by noting that $d \underline{\Sigma}_\mu \, \delta x^\mu$ corresponds to the volume swept out by the variation of the boundary induced by the displacement $\delta x^\mu$. The surface element $d \bar{\Sigma}_\mu$ is defined as
\begin{equation} \label{WAC-VolumeElement}
d \underline{\Sigma}_\mu := \frac{1}{3!} \epsilon_{\mu \alpha \beta \gamma } \, dx^\alpha \wedge dx^\beta \wedge dx^\gamma.
\end{equation}

\noindent One can (by way of the divergence theorem) convert \ref{WAC-BulkIntVar0} to a bulk integral of the form:
\begin{equation} \label{WAC-BulkIntVar}
\delta I = \int_U d^4 x \left[ \delta F + \partial_\mu (F \, \delta x^\mu) \right].
\end{equation}

\noindent Defining
\begin{equation} \label{WAC-functiondensity}
f:=F/\sqrt{-g} ,
\end{equation}

\noindent one may write:
\begin{equation} \label{WAC-BulkIntVarCov0}
\delta I = \int_U d^4 x \left[ \sqrt{-g} \left(\delta f - \frac{1}{2} g_{\mu \nu} \delta g^{\mu \nu} f \right) + \partial_\mu (\sqrt{-g} f \delta x^\mu) \right].
\end{equation}

\noindent If $f$ is a scalar, one may write this as:
\begin{equation} \label{WAC-BulkIntVarCov1}
\begin{aligned}
\delta I &= \int_U d^4 x \, \sqrt{-g} \left[\delta f - \frac{1}{2} g_{\mu \nu} \delta g^{\mu \nu} f + \bar{\nabla}_\mu (f \delta x^\mu) \right] \\
&= \int_U d^4 x \, \sqrt{-g} \left[\delta f - \frac{1}{2} g_{\mu \nu} \delta g^{\mu \nu} f \right] + \int_{\partial U} d \Sigma_\mu \, \delta x^\mu \, f .
\end{aligned}
\end{equation}

\noindent where $d \Sigma_\mu := \sqrt{-g} \, d \underline{\Sigma}_\mu $.

\section{Variation of covariant divergence}\label{Appcovdiv}

Here, we consider the variation of a boundary integral of the following form:
\begin{equation} \label{WAC-BdyInt}
B = \int_U d^4 x  \sqrt{-g} \left( {\nabla}_\mu V^\mu \right) ,
\end{equation}

\noindent where $V^\mu$ are the components of some vector field. First, we consider the variation of the divergence itself:
\begin{equation} \label{WAC-BdyIntgrandVar}
\begin{aligned}
\delta [ {\nabla}_\mu V^\mu ]
     &= {\nabla}_\mu \delta V^\mu + V^\nu \delta {\Gamma}{^\mu}_{\nu \mu} \\
%
%
%
     &= {\nabla}_\mu \delta V^\mu + \frac{1}{2}V^\nu [ g^{\mu \sigma} {\nabla}_\nu \delta g_{\sigma \mu} ] \\
%
%
&= {\nabla}_\sigma \delta V^\sigma + \frac{1}{2} g_{\mu \nu} [{\nabla}_\sigma V^\sigma ] \delta g^{\mu \nu} \\
& \quad - \frac{1}{2} g_{\mu \nu} {\nabla}_\sigma [ V^\sigma \delta g^{\mu \nu}]
.
\end{aligned}
\end{equation}

\noindent This result may be combined with the variation of the volume element to obtain:
\begin{equation} \label{WAC-BdyIntVar1}
\begin{aligned}
\delta B =& \int_U d^4 x  \sqrt{-g} \, {\nabla}_\sigma \left[ \delta V^\sigma - \frac{1}{2} \, V^\sigma \, g_{\mu \nu} \, \delta g^{\mu \nu} \right] \\
& + \int_{\partial U} \, d\Sigma_\mu \delta x^\mu \, {\nabla}_\nu V^\nu,
\end{aligned}
\end{equation}

\noindent which can be rewritten:
\begin{equation} \label{WAC-BdyIntVar}
\begin{aligned}
\delta B =& \int_U d^4 x  \sqrt{-g} \, {\nabla}_\sigma \left[ \delta V^\sigma - \frac{1}{2} \, V^\sigma \, g_{\mu \nu} \, \delta g^{\mu \nu} \right] \\
& + \int_{\partial U} \, i_{\delta x} \left[ {\nabla}_\nu V^\nu \, \omega \right],
\end{aligned}
\end{equation}

\noindent where the last term comes from the dispacement $\delta x^\mu=\Delta \lambda \, \xi^\mu$ of the boundary, where the volume form $\omega$ is defined:
\begin{equation} \label{WAC-VolumeForm}
\omega := \frac{\sqrt{-g}}{4!} \, \epsilon_{\alpha \beta \mu \nu} \, dx^\alpha \wedge dx^\beta \wedge dx^\mu \wedge dx^\nu.
\end{equation}

\noindent The interior product of the volume form is:
\begin{equation} \label{WAC-VolumeFormIntProd}
i_{V}\omega := \frac{\sqrt{-g}}{3!} \, \epsilon_{\alpha \beta \mu \nu} \, V^\alpha \, dx^\beta \wedge dx^\mu \wedge dx^\nu = d\Sigma_\sigma \, V^\sigma.
\end{equation}

\noindent Now write:
\begin{equation} \label{WAC-BdyVar}
\begin{aligned}
\Delta V^\sigma &= \delta V^\sigma + \Delta \lambda \, \pounds_\xi V^\sigma \\
\Delta g^{\mu \nu} &= \delta g^{\mu \nu} + \Delta \lambda \, \pounds_\xi g^{\mu \nu} ,
\end{aligned}
\end{equation}

\noindent which yields:
\begin{equation} \label{WAC-BdyIntVar2}
\begin{aligned}
\delta B =& \int_{\partial U} d\Sigma_\sigma \left[ \Delta V^\sigma - \frac{1}{2} \, V^\sigma \, g_{\mu \nu} \, \Delta g^{\mu \nu} \right] \\
& - \Delta \lambda \int_{\partial U} \, d\Sigma_\sigma \left[\pounds_\xi V^\sigma - \frac{1}{2} \, V^\sigma \, g_{\mu \nu} \, \pounds_\xi g^{\mu \nu} \right] \\
& + \int_{\partial U} \, i_{\delta x} \left[ {\nabla}_\nu V^\nu \, \omega \right].
\end{aligned}
\end{equation}

\noindent This can be written as (using the identity $d i_V \omega=\nabla_\mu V^\mu \omega$):
\begin{equation} \label{WAC-BdyIntVar3}
\begin{aligned}
\delta B =& \int_{\partial U} d\Sigma_\sigma \left[ \Delta V^\sigma - \frac{1}{2} \, V^\sigma \, g_{\mu \nu} \, \Delta g^{\mu \nu} \right] \\
& - \Delta \lambda \int_{\partial U} \, \pounds_\xi i_V \omega  + \Delta \lambda \int_{\partial U} \, i_{\xi} d i_V \omega .
\end{aligned}
\end{equation}

\noindent From Cartan's magic formula $di+id=\pounds$, one can write:
\begin{equation} \label{WAC-Magic}
\pounds_\xi i_V \omega = d i_\xi i_V \omega +  i_\xi d i_V \omega.
\end{equation}

\noindent After a cancellation, one finds that:
\begin{equation} \label{WAC-BdyIntVar4a}
  \begin{aligned}
  \delta B =& \int_{\partial U} d\Sigma_\sigma \left[ \Delta V^\sigma - \frac{1}{2} V^\sigma  g_{\mu \nu} \Delta g^{\mu \nu} \right]  - \Delta \lambda \int_{\partial U} d i_\xi i_V \omega .
  \end{aligned}
\end{equation}

\noindent The term proportional to $\Delta \lambda$ on the right-hand side of \ref{WAC-BdyIntVar4a}, being the exterior derivative of a (two-)form, vanishes when integrated over $\partial U$ via the boundary of a boundary principle, but we choose to keep it for now. The second interior product takes the form:
\begin{equation} \label{WAC-VolumeFormIntProd2}
  i_{\xi}i_{V}\omega := \frac{\sqrt{-g}}{2!} \, \epsilon_{\alpha \beta \mu \nu} \, V^\alpha  \, \xi^\beta \wedge dx^\mu \wedge dx^\nu .
\end{equation}

\noindent The exterior derivative takes the form:
\begin{equation} \label{WAC-VolumeFormdIntProd2Exterior}
  \begin{aligned}
  di_{\xi}i_{V}\omega
    &:= \frac{1}{3!} \partial_\sigma \left(\sqrt{-g} \, \epsilon_{\alpha \beta \mu \nu} \, V^\alpha \, \xi^\beta \right)  dx^\sigma \wedge dx^\mu \wedge dx^\nu \\
    &\,= \frac{1}{3!}  \, \partial_\sigma \left(\sqrt{-g} \, \epsilon_{\alpha \beta \mu \nu} \, V^{[\alpha} \, \xi^{\beta]}\right)  dx^\sigma \wedge dx^\mu \wedge dx^\nu .
\end{aligned}
\end{equation}

\noindent We recognize the quantity in the bracket as the Hodge star of the antisymmetric tensor $V^{[\alpha} \, \xi^{\beta]}$. Now consider the following expression for the antisymmetric tensor $A^{\mu \nu}$:
\begin{equation} \label{WAC-dHAntisym1}
  \begin{aligned}
  d{^\star}A
    &:= \frac{1}{3!} \partial_{[\sigma} \left({^\star}A_{\mu \nu]} \right) dx^\sigma \wedge dx^\mu \wedge dx^\nu ,
\end{aligned}
\end{equation}

\noindent where ${^\star}A_{\mu \nu} = \sqrt{-g} \epsilon{_{\mu \nu \alpha \beta}} \, A^{\alpha \beta}$ are the components of the Hodge dual of the tensor $A^{\mu \nu}$. In a four-dimensional Lorentzian manifold, the double dual operator of a $3$-form is the identity, so one may write:
\begin{equation} \label{WAC-dHAntisym2}
  \begin{aligned}
  d{^\star}A
    &= \frac{1}{3!} \epsilon_{\sigma \mu \nu \tau} \, \epsilon^{\tau \alpha \beta \gamma} \partial_{[\alpha} \left({^\star}A_{\beta \gamma]} \right) dx^\sigma \wedge dx^\mu \wedge dx^\nu \\
    &= - \epsilon^{\tau \alpha \beta \gamma} \partial_{\alpha} \left({^\star}A_{\beta \gamma} \right) d\underline{\Sigma}_\tau \\
    &= 2 \partial_{\alpha} \left(\sqrt{-g} A^{\tau \alpha} \right) d\underline{\Sigma}_\tau
\end{aligned}
\end{equation}

\noindent It follows that the expression in \ref{WAC-VolumeFormdIntProd2Exterior} can be rewritten:
\begin{equation} \label{WAC-VolumeFormdIntProd2Exterior2}
  \begin{aligned}
  di_{\xi}i_{V}\omega
    &:= 2 \partial_{\alpha} \left(\sqrt{-g} V^{[\sigma} \, \xi^{\alpha]}\right) d\underline{\Sigma}_\sigma,
\end{aligned}
\end{equation}

\noindent so that:
\begin{equation} \label{WAC-BdyIntVar4}
  \begin{aligned}
  \delta B =& \int_{\partial U} d\Sigma_\sigma \left[ \Delta V^\sigma - \frac{1}{2} V^\sigma  g_{\mu \nu} \Delta g^{\mu \nu} \right] \\
  & - 2 \Delta \lambda \int_{\partial U} \nabla_{\alpha} \left(V^{[\sigma} \, \xi^{\alpha]}\right) d{\Sigma}_\sigma .
  \end{aligned}
\end{equation}

\section{Lie derivative of connection coefficients} \label{APPX-LDConn}

The Lie derivative measures the deviation from formal invariance under a diffeomorphism. We consider here a one-parameter diffeomorphism $x^\prime(\epsilon,x)$, which to first order in $\epsilon$, takes the form:
\begin{equation} \label{WAC-DiffeomorphismExpansion}
x^\prime(\epsilon,x) = x + \epsilon \, \xi(x) + O(\epsilon^2).
\end{equation}

\noindent To first order in $\epsilon$, this can be easily inverted to obtain (noting that $\xi(x^\prime)=\xi(x) + O(\epsilon)$):
\begin{equation} \label{WAC-DiffeomorphismExpansionInvert}
x(\epsilon,x^\prime) = x^\prime - \epsilon \, \xi(x^\prime) + O(\epsilon^2),
\end{equation}

\noindent and it follows that:
\begin{equation} \label{WAC-LieDerivativeMetric}
\begin{aligned}
\frac{\partial {x^\prime}^\mu}{\partial x^\nu}
&= \delta{^\mu}{_\nu} + \epsilon \, \frac{\partial \xi^\mu}{\partial x^\nu} + O(\epsilon^2) \\
\frac{\partial x^\mu}{\partial {x^\prime}^\nu}
&= \delta{^\mu}{_\nu} - \epsilon \, \frac{\partial \xi^\mu}{\partial {x^\prime}^\nu} + O(\epsilon^2) .
\end{aligned}
\end{equation}

\noindent For a tensor field $T^{\mu...}_{\nu...}(x)$, one can characterize the deviation from formal invariance under the diffeomorphism $x^\prime(\epsilon,x)$ in the following way:
\begin{equation} \label{WAC-LieDerivDefn}
(\pounds_\xi T)^{\mu...}_{\nu...}(x):=\lim_{\epsilon \to 0} \frac{1}{\epsilon}\left\{T^{\mu...}_{\nu...}(x^\prime(\epsilon,x))-{T^\prime}^{\mu...}_{\nu...} (x^\prime(\epsilon,x))\right\},
\end{equation}

\noindent where ${T^\prime}^{\mu...}_{\nu...}$ represents the components of the coordinate transformed tensor. The above formula defines the Lie derivative and may be applied directly to the connection coefficients.

To work out the explicit expression for the Lie derivative, first recall the transformation law for connection coefficients:
  \begin{equation} \label{WAC-ConnectionCoeffsTransformation}
  {\bar{\Gamma}^\prime}{^\gamma}{_{\mu \nu}}=\left[\frac{\partial {x^\prime}^\gamma}{\partial x^\sigma}\frac{\partial x^\alpha}{\partial {x^\prime}^\mu}\frac{\partial x^\beta}{\partial {x^\prime}^\nu}\right]\bar{\Gamma}^\sigma{_{\alpha \beta}}
  -
  \frac{\partial x^\alpha}{\partial {x^\prime}^\mu}\frac{\partial x^\beta}{\partial {x^\prime}^\nu} \left[\frac{\partial^2 {x^\prime}^\gamma}{\partial x^\alpha \partial x^\beta}\right].
  \end{equation}

\noindent Expanding ${\bar{\Gamma}^\prime}{^\gamma}{_{\mu \nu}}(x^\prime(\epsilon,x))$ to lowest order in $\epsilon$ (quantities without arguments are implicitly functions of $x$),
\begin{equation} \label{WAC-ConnectionCoeffsExp}
\begin{aligned}
{\bar{\Gamma}^\prime}{^\gamma}{_{\mu \nu}}(x^\prime(\epsilon,x))
=&~ {\bar{\Gamma}^\prime}{^\gamma}{_{\mu \nu}}
 + \epsilon \, \xi^\sigma \frac{\partial{\bar{\Gamma}^\prime}{^\gamma}{_{\mu \nu}}}{\partial x^\sigma} + O(\epsilon^2),
\end{aligned}
\end{equation}

\noindent and expanding ${\bar{\Gamma}^\prime}{^\gamma}{_{\mu \nu}}(x^\prime(\epsilon,x))$ to lowest order in $\epsilon$, we obtain the following expression:
\begin{equation} \label{WAC-ConnectionCoeffsTransformedExp}
\begin{aligned}
{\bar{\Gamma}^\prime}{^\gamma}{_{\mu \nu}}(x^\prime(\epsilon,x))
=&~ {\bar{\Gamma}^\prime}{^\gamma}{_{\mu \nu}}
 + \epsilon \frac{\partial \xi^\gamma}{\partial x^\sigma} \bar{\Gamma}^\sigma{_{\mu \nu}}
 - \epsilon \frac{\partial \xi^\sigma}{\partial x^\mu} {\bar{\Gamma}^\prime}{^\gamma}{_{\sigma \nu}} \\
& - \epsilon \frac{\partial \xi^\sigma}{\partial x^\nu} {\bar{\Gamma}^\prime}{^\gamma}{_{\mu \sigma}} - \epsilon \frac{\partial^2 \xi^\gamma}{\partial x^\mu \partial x^\nu} + O(\epsilon^2).
\end{aligned}
\end{equation}

\noindent The Lie derivative of the connection coefficients then takes the form:
\begin{equation} \label{WAC-ConnectionCoeffsLD}
\begin{aligned}
\pounds_\xi {\bar{\Gamma}^\prime}{^\gamma}{_{\mu \nu}}
=&~ \xi^\sigma \frac{\partial{\bar{\Gamma}^\prime}{^\gamma}{_{\mu \nu}}}{\partial x^\sigma}
- \frac{\partial \xi^\gamma}{\partial x^\sigma} \bar{\Gamma}^\sigma{_{\mu \nu}}
+ \frac{\partial \xi^\sigma}{\partial x^\mu} {\bar{\Gamma}^\prime}{^\gamma}{_{\sigma \nu}} \\
& + \frac{\partial \xi^\sigma}{\partial x^\nu} {\bar{\Gamma}^\prime}{^\gamma}{_{\mu \sigma}} + \frac{\partial^2 \xi^\gamma}{\partial x^\mu \partial x^\nu}.
\end{aligned}
\end{equation}

\noindent Now one might expect $\pounds_\xi {\bar{\Gamma}^\prime}{^\gamma}{_{\mu \nu}}$ to be covariant. To see that it is in fact covariant, consider the following expression:
\begin{equation} \label{WAC-CDxi}
\begin{aligned}
\frac{\partial \xi^\mu}{\partial x^\nu} = \bar{\nabla}_\nu \xi^\mu - \bar{\Gamma}^\mu{_{\nu \tau}} \xi^\tau,
\end{aligned}
\end{equation}

\noindent which is just a rewriting of the formula for the covariant derivative $\bar{\nabla}_\nu \xi^\mu$.
One can show that:
\begin{equation} \label{WAC-ConnectionCoeffsLD2c}
\begin{aligned}
\pounds_\xi {\bar{\Gamma}^\prime}{^\gamma}{_{\mu \nu}}
=&~ \xi^\sigma [ \partial_\sigma {\bar{\Gamma}^\prime}{^\gamma}{_{\mu \nu}} - \partial_\mu {\bar{\Gamma}^\prime}{^\gamma}{_{\nu \sigma}}
 +  \bar{\Gamma}^\tau{_{\mu \nu}} {\bar{\Gamma}^\prime}{^\gamma}{_{\tau \sigma}} \\
& + {\bar{\Gamma}^\prime}{^\gamma}{_{\nu \tau}} \bar{\Gamma}^\tau{_{\mu \sigma}} - {\bar{\Gamma}^\prime}{^\gamma}{_{\tau \nu}} \bar{\Gamma}^\tau{_{\mu \sigma}} - {\bar{\Gamma}^\prime}{^\gamma}{_{\mu \tau}} \bar{\Gamma}^\tau{_{\nu \sigma}} ]\\
& + \bar{\nabla}_\mu (\bar{\nabla}_\nu \xi^\gamma) + \bar{\nabla}_\mu \xi^\sigma T^\gamma{_{\sigma \nu}},
\end{aligned}
\end{equation}

\noindent where the torsion tensor is $ T^\gamma{_{\mu \nu}}:= {\bar{\Gamma}^\prime}{^\gamma}{_{\mu \nu}} - {\bar{\Gamma}^\prime}{^\gamma}{_{\nu \mu}}$. One may write:
\begin{equation} \label{WAC-TorsionRelation}
\begin{aligned}
{\bar{\Gamma}^\prime}{^\gamma}{_{\mu \nu}} = {\bar{\Gamma}^\prime}{^\gamma}{_{\nu \mu}} + T^\gamma{_{\mu \nu}}.
\end{aligned}
\end{equation}
    %

\noindent One can use the above to obtain:
\begin{equation} \label{WAC-ConnectionCoeffsLD3b}
\begin{aligned}
\pounds_\xi {\bar{\Gamma}^\prime}{^\gamma}{_{\mu \nu}}
=&~ \xi^\sigma [ \partial_\sigma {\bar{\Gamma}^\prime}{^\gamma}{_{\mu \nu}} - \partial_\mu {\bar{\Gamma}^\prime}{^\gamma}{_{\sigma \nu}}
 + \bar{\Gamma}^\tau{_{\mu \nu}} {\bar{\Gamma}^\prime}{^\gamma}{_{\sigma \tau}} - {\bar{\Gamma}^\prime}{^\gamma}{_{\mu \tau}} \bar{\Gamma}^\tau{_{\sigma \nu}}\\
& - \partial_\mu T^\gamma{_{\nu \sigma}} + \bar{\Gamma}^\tau{_{\mu \nu}} T^\gamma{_{\tau \sigma}} + \bar{\Gamma}^\tau{_{\mu \sigma}}  T^\gamma{_{\nu \tau}} \\
& - {\bar{\Gamma}^\prime}{^\gamma}{_{\mu \tau}} T^\tau{_{\sigma \nu}}] + \bar{\nabla}_\mu (\bar{\nabla}_\nu \xi^\gamma) + \bar{\nabla}_\mu \xi^\sigma T^\gamma{_{\sigma \nu}},
\end{aligned}
\end{equation}
    %

\noindent and upon identifying the curvature and covariant derivative of the torsion tensor, one obtains the result:
\begin{equation} \label{WAC-ConnectionCoeffsLD3d}
\begin{aligned}
\pounds_\xi {\bar{\Gamma}^\prime}{^\gamma}{_{\mu \nu}}
=&~ \bar{\nabla}_\mu \bar{\nabla}_\nu \xi^\gamma + \xi^\sigma \bar{R}{^\gamma}_{\nu \sigma \mu} + \bar{\nabla}_\mu \left(\xi^\sigma T^\gamma{_{\sigma \nu}} \right).
\end{aligned}
\end{equation}

\section{Divergence expression}
Here, we compare the divergences of a tensor with respect to two different metric-compatible connections. First, a useful result is derived for two metric tensors $g_{\mu \nu}$ and $\eta_{\mu \nu}$:
\begin{equation}\label{GREC-PDGam}
  \begin{aligned}
  \partial_\sigma \left[\frac{\sqrt{-\eta}}{\sqrt{-g}}\right]
  & =
  \frac{\sqrt{-\eta}}{\sqrt{-g}}
  \left[
    \frac{\partial_\sigma \sqrt{-\eta}}{\sqrt{-\eta}}
    -
    \frac{\partial_\sigma \sqrt{-g}}{\sqrt{-g}}
  \right] \\
  & =
  - \frac{\sqrt{-\eta}}{\sqrt{-g}}
  \left[
    \Gamma{^\tau}_{\tau \sigma}
    -
    \bar{\Gamma}{^\tau}_{\tau \sigma}
  \right] =
  - \frac{\sqrt{-\eta}}{\sqrt{-g}} \, W{^\tau}_{\tau \sigma} .
  \end{aligned}
\end{equation}

\noindent For a tensor $T{^\mu}{_\nu}$ and a scalar field $\varphi$, one may obtain the following expression (where $\bar{\nabla}_\alpha \eta_{\mu \nu} = 0$):
\begin{equation} \label{GREC-dT}
  \begin{aligned}
  \nabla_\mu (\varphi \, T{^\mu}{_\nu})
  = \, & \bar{\nabla}_\mu (\varphi \, T{^\mu}{_\nu}) + \varphi \, W{^\mu}_{\mu \sigma} \, T{^\sigma}{_\nu} - \varphi \, W{^\sigma}_{\mu \nu} \, T{^\mu}{_\sigma} \\
  = \, & \varphi \, \bar{\nabla}_\mu T{^\mu}{_\nu} + \varphi \, W{^\mu}_{\mu \sigma} \, T{^\sigma}{_\nu} - \varphi \, W{^\sigma}_{\mu \nu} \, T{^\mu}{_\sigma} \\
  & + T{^\mu}{_\nu} \, \partial_\mu \varphi .
  \end{aligned}
\end{equation}

\noindent Now if $\varphi={\sqrt{-\eta}}/{\sqrt{-g}}$, one can show that:
\begin{equation} \label{GREC-dT2}
  \begin{aligned}
  {\sqrt{-g}} \, \nabla_\mu \left[\frac{\sqrt{-\eta}}{\sqrt{-g}} \, T{^\mu}{_\nu}\right]
  = \, &
  \sqrt{-\eta} \left( \bar{\nabla}_\mu T{^\mu}{_\nu} - W{^\sigma}_{\mu \nu} \, T{^\mu}{_\sigma} \right).
  \end{aligned}
\end{equation}

We now consider the divergence of a current $k^\mu = \xi^\tau \, \tilde{T}{^\mu}{_\tau}$, where $\xi^\mu$ is the dual of a vector field:
\begin{equation}\label{GREC-DivXiT}
  \begin{aligned}
  \nabla_\sigma k^\sigma
  = \, &
  \nabla_\sigma \left(\xi^\tau \, \tilde{T}{^\sigma}{_\tau}\right) \\
  = \, &
  \xi^\tau \nabla_\sigma \tilde{T}{^\sigma}{_\tau}
  + \tilde{T}{^\sigma}{_\tau} \, \nabla_\sigma \xi^\tau \\
  = \, &
  \xi^\tau \nabla_\sigma \tilde{T}{^\sigma}{_\tau}
  +
  \tilde{T}{^\sigma}{_\tau} \, \bar{\nabla}_\sigma \xi^\tau
  +
  \tilde{T}{^\sigma}{_\nu} \, W{^\nu}_{\sigma \tau} \, \xi^\tau\\
  = \, &
  \left(
  \nabla_\sigma \tilde{T}{^\sigma}{_\tau}
  +
  \tilde{T}{^\sigma}{_\nu} \, W{^\nu}_{\sigma \tau}
  \right) \xi^\tau\
  +
  \tilde{T}{^\sigma}{_\tau} \, \bar{\nabla}_\sigma \xi^\tau
  .
  \end{aligned}
\end{equation}

\noindent Upon setting $\tilde{T}^{\mu \nu} = \varphi \, T^{\mu \nu}$, and substituting \ref{GREC-dT}:
\begin{equation}\label{GREC-DivXiT2}
  \begin{aligned}
  \nabla_\sigma (\varphi \, \xi^\tau \, {T}{^\sigma}{_\tau})
  = \, &
  \xi^\tau \left[
  \nabla_\sigma (\varphi \, {T}{^\sigma}{_\tau} )
  +
  \varphi \, {T}{^\sigma}{_\nu} \, W{^\nu}_{\sigma \tau}
  \right] \\
  &
  +
  \varphi \, {T}{^\sigma}{_\tau} \, \bar{\nabla}_\sigma \xi^\tau \\
  = \, &
  \xi^\tau \biggl[
  \varphi \, \bar{\nabla}_\sigma {T}{^\sigma}{_\tau}
  +
  T{^\mu}{_\tau} \, \partial_\mu \varphi
  +
  \varphi \, W{^\sigma}_{\sigma \mu} \, T{^\mu}{_\tau}
  \biggr]\\
  & \quad
  +
  \varphi \, {T}{^\sigma}{_\tau} \, \bar{\nabla}_\sigma \xi^\tau
  ,
  \end{aligned}
\end{equation}

\noindent and setting $\varphi={\sqrt{-\eta}}/{\sqrt{-g}}$, we obtain the following result:
\begin{equation}\label{GREC-DivXiT3}
  \begin{aligned}
  \nabla_\sigma
  \left[
    \frac{\sqrt{-\eta}}{\sqrt{-g}} \, \xi^\tau \, {T}{^\sigma}{_\tau}
  \right]
  =
  \frac{\sqrt{-\eta}}{\sqrt{-g}} \biggl[
    &
    \bar{\nabla}_\sigma {T}{^\sigma}{_\tau} \, \xi^\tau
    +
    {T}{^\sigma}{_\tau} \, \bar{\nabla}_\sigma \xi^\tau \biggr].
  \end{aligned}
\end{equation}

\noindent Alternatively, defining the tensor $S^{\sigma \mu} = \varphi \, T^{\sigma \mu}$, the above may be rewritten:
\begin{equation}\label{GREC-DivXiT3B}
  \begin{aligned}
  \nabla_\sigma
  \left[
    \xi^\tau \, {S}{^\sigma}{_\tau}
  \right]
  = \, &
    \frac{\sqrt{-\eta}}{\sqrt{-g}} \,
    \xi^\tau \,
    \bar{\nabla}_\sigma
    \left(
      \frac{\sqrt{-g}} {\sqrt{-\eta}}
      {S}{^\sigma}{_\tau}
    \right)
    +
    {S}{^\sigma}{_\tau} \, \bar{\nabla}_\sigma \xi^\tau
  .
  \end{aligned}
\end{equation}


%
%


\bibliography{Weiss}

\end{document}